\documentclass[aps,prb, twocolumn,floatfix,notitlepage, superscriptaddress,longbibliography]{revtex4-2}
\usepackage[utf8]{inputenc}
\usepackage{amsfonts,amsmath,amssymb,graphicx,epstopdf,verbatim}
\usepackage{rotating}
\usepackage{epstopdf}
\usepackage{xcolor}
\usepackage{natbib}
\usepackage[colorlinks]{hyperref}
\usepackage{overpic}
\usepackage{dsfont}
\usepackage{comment}
\definecolor{darkred}{rgb}{0.5,0,0}
\definecolor{darkgreen}{rgb}{0,0.5,0}
\definecolor{darkblue}{rgb}{0,0,0.5}
\hypersetup{colorlinks,
linkcolor=darkred,
filecolor=darkgreen,
urlcolor=darkblue,
citecolor=darkgreen}
\usepackage{braket}
\usepackage{overpic}

\newcommand{\calN}{\mathcal{N}}

\newcommand{\nep}{e}

\newcommand{\hK}{\hat{K}}
\newcommand{\hS}{\hat{S}}
\newcommand{\hSw}{\hat{P}}
\renewcommand{\H}{\hat{H}}
\newcommand{\hU}{\hat{U}}
\newcommand{\hsigma}{\hat{\sigma}}
\newcommand{\MBL}{\text{int}}
\newcommand{\vsigma}{\vec{\boldsymbol{\sigma}}}
\newcommand{\Id}{1}
\newcommand{\vS}{\vec{\mathbf{S}}}

\newcommand{\hLI}{\hat{\mathcal{I}}_{\rm LI}}
\newcommand{\LI}{\mathcal{I}_{\rm LI}}
\newcommand{\abs}[1]{|#1|}
\newcommand{\sign}{\operatorname{sign}}
\newcommand{\AF}{\mathrm{AF}}

\newcommand{\calI}{\mathcal{I}}

\renewcommand{\i}{\boldsymbol{i}}

\begin{document}

\title{Swapping Floquet time crystal}

\author{Roberto Gargiulo}
\email{r.gargiulo@fz-juelich.de}
\altaffiliation{Present affiliation: Forschungszentrum Jülich GmbH, Peter Gr\"unberg Institute, Quantum Control (PGI-8), 52425 J\"ulich, Germany}
\affiliation{Dipartimento di Fisica ``E. Pancini", Universit\`a di Napoli ``Federico II'', Monte S. Angelo, I-80126 Napoli, Italy}

\author{Gianluca Passarelli}
\affiliation{Dipartimento di Fisica ``E. Pancini", Universit\`a di Napoli ``Federico II'', Monte S. Angelo, I-80126 Napoli, Italy}

\author{Procolo Lucignano}
\affiliation{Dipartimento di Fisica ``E. Pancini", Universit\`a di Napoli ``Federico II'', Monte S. Angelo, I-80126 Napoli, Italy}

\author{Angelo Russomanno}
\altaffiliation{Present affiliation: Dipartimento di Fisica ``E. Pancini", Universit\`a di Napoli ``Federico II'', Monte S. Angelo, I-80126 Napoli, Italy}
\affiliation{Scuola Superiore Meridionale, Università di Napoli Federico II, Largo San Marcellino 10, I-80138 Napoli, Italy}

\begin{abstract}
 We propose a Floquet period-doubling time-crystal model based on a disordered interacting long-range spin chain where the periodic swapping of nearby spin couples is applied. This protocol can be applied to systems with any local spin magnitude $s$ {and in principle also to systems with nonspin (fermionic or bosonic) local Hilbert space}. We explicitly consider the cases $s = 1/2$ and $s = 1$, using analytical and numerical methods to show that the time-crystal behavior appears in a range of parameters. In particular, we study the persistence of period-doubling oscillations in time, the time-crystal properties of the Floquet spectrum (quasienergy $\pi$-spectral pairing and long-range correlations of the Floquet states), and introduce a quantity (the local imbalance) to assess what initial states give rise to a period-doubling dynamics. We also consider the average level spacing ratio and find that the interval of parameters where the system does not thermalize and persistent period-doubling is possible corresponds to the one where the Floquet spectrum shows time-crystal properties.
%
\end{abstract}

\maketitle

\section{Introduction}

The experimental discovery~\cite{Zhang2017,Choi2017} of time-crystals few years after their theoretical prediction~\cite{Wilczek2012,shapere2012classical,wilczek2013superfluidity} has been a real breakthrough. In analogy to ordinary crystals, time crystals appear as a consequence of breaking  time-translation symmetry in the system~\cite{khemani2019brief,sacha2017time,annurev-conmatphys,RevModPhys.95.031001,sacha2020time}. Following earlier attempts to identify systems able to display time-translation symmetry breaking, a no-go theorem showed that this is not possible in the ground state or in thermal equilibrium of physical Hamiltonians~\cite{watanabe2015absence,noter}.

Among many possible non-equilibrium candidates, quantum periodically-driven (Floquet) systems have proven to be the most promising realization. Here the discrete time-translation symmetry is broken in the thermodynamic limit, by the appearance of a response with a period multiple than the one of the driving. Stimulated by
the initial proposals~\cite{Nayak2016,Khemani2016}, a large body of theoretical work has been performed~\cite{PhysRevB.96.115127,Potter2017,PhysRevLett.119.010602,pizzi2020time,smits2018observation,pizzi2019period,
pizzi2021higher,PhysRevB.95.214307,Surace_2019,zhu2019dicke,lazarides2017fate,pizzi2021bistability,
gong2018discrete,else2017prethermal,Reyhaneh,note_tct}. A common ingredient is the presence in the dynamics of a sufficient number of
constraints that introduce ergodicity-breaking, thus preventing infinite-temperature thermalization, for instance disorder that induces many-body localization (MBL); see~\cite{RevModPhys.91.021001} for a review.

In this context, the properties of the Floquet states~\cite{Sambe1973,shirley1965solution} -- the eigenstates of the periodically-driven dynamics -- are crucial. In order for a period-doubling Floquet time crystal to occur, all of the Floquet states must be organized in doublets of cat states. {Cat states are superpositions of two macroscopically different classical spin configurations, and} the time crystal appears as a form of Rabi oscillation between th{ese configurations}, with frequency provided by the splitting of the doublet. In the thermodynamic limit, the oscillation frequencies of all the doublets become synchronized, and this synchronization occurs at period double than the driving ($\pi$-spectral pairing), resulting in a time crystal. In the original proposal all the Floquet spectrum was $\pi$-spectral paired, but there are also works where only a fraction of the Floquet spectrum is paired, and a period-doubling behavior for specific initial states can be observed~\cite{huang2023analytical,PhysRevLett.129.133001,PhysRevLett.120.110603}.

Following these results, we propose a periodic-driving protocol, which is applied to a disordered spin chain with long-range interactions, and leads to a Floquet time crystal. The drive consists in periodically \textit{swapping} two neighbouring spins, instead of flipping each individual spin (see Fig.~\ref{fig:Comparison_Flip_Swap}). As in the spin-flipping case, this protocol can be implemented by periodically applying in sequence two different Hamiltonians, and can be applied to interacting chains of particles with arbitrary value of the local spin magnitude $s$, rather than only to spin-1/2 systems. {Moreover,  the swap protocol has a straightforward physical meaning in systems with fermionic or bosonic local Hilbert space, since it conserves particle number. Therefore it has a potential range of applicability that goes beyond spin systems}. 

We find that a Floquet time crystal is realized in this case, although in a slightly weaker sense than what has been found for the spin-flip driving. While in Refs.~\cite{Nayak2016,Khemani2016} all the Floquet spectrum is $\pi$-spectral paired, here a fraction of the quasienergies does not have this property. Nevertheless, this fraction tends exponentially to zero with the system size, and we can observe persisting and robust period-doubling oscillations for macroscopically many initial states.

The paper is organized as follows. In Sec.~\ref{sec:model} we introduce our model Hamiltonians -- both for local spins with magnitude 1/2 and for generic magnitudes -- and study how the period doubling appears in the solvable case, a special point in the parameter space where analytical computations can be made. 

In Sec.~\ref{li:sec} we introduce the local imbalance, a quantity that allows characterizing the initial classical spin configurations that give rise to a period-doubling dynamics. We find that increasing the local spin magnitude, the distribution of the imbalances on the spin configurations gets a larger average and becomes narrower, marking that a period-doubling behavior occurs more easily for a random initial spin configuration.

In Sec.~\ref{peridolo:sec} we start with our numerical analysis, that we perform for the models with local spin  $s=1/2$ (spin-1/2 case) and $s = 1$ (spin-1 case). The numerical analysis is important in order to assess if the period-doubling behavior is robust and persists also beyond the solvable point, which would indicate a proper time-crystal phase supported by the interactions, rather than an isolated point. We study the period doubling in the time domain, looking at a collective quantity that is nonvanishing whenever period-doubling oscillations of the local magnetizations are present.

 Moving away from the solvable point, period doubling oscillations are still there, but last a finite time that exponentially increases with the system size. In this way period-doubling oscillations become persistent in the large-size limit, as appropriate for a time crystal, for which the time-translation symmetry breaking appears only in the thermodynamic limit, in analogy with standard symmetry breaking~\cite{sachdev2011quantum}. In doing our analysis, we emphasize that for some initial states only part of the onsite magnetizations can show period doubling. 

In order to see a nontrivial period-doubling behavior of the observables, there must be no energy absorption and no infinite-temperature thermalization (in this sense the dynamics must be regular or integrable-like). In order to find the range of parameters where this occurs, we focus in Sec.~\ref{regdyn:sec} on the average level-spacing ratio, a standard probe of the integrability/ergodicity properties of the quantum dynamics~\cite{Pal_PhysRevB10,PhysRevLett.110.084101}. We find an interval of the parameters where the dynamics is integrable-like, and see that this effect becomes more marked with increasing system size. {In this interval there is no thermalization and we can see the} period-doubling oscillations. We argue that the physical effect leading to this breaking of ergodicity is disorder, inducing MBL.

In Sec.~\ref{floq:sec} we study the $\pi$-spectral-pairing properties of the Floquet states, and see that this phenomenon is strictly related to period doubling. The $\pi$-spectral pairing can be found analytically at the {solvable} point, and we numerically find that it is robust in a well-defined range of parameters that marks the time-crystal phase. This robustness is related to the disordered long-range interactions that break the degeneracies in the Floquet spectrum at the solvable point, and make the structure of Floquet states and quasienergies robust under local perturbations (outside of the solvable point, the $\pi$-spectral pairing appears rigorously only in the thermodynamic limit. At finite size there is an error leading to the finite-time duration of the period-doubling oscillations -- see Appendix~\ref{app:pidoub}). 

In this section we also directly study the cat-state properties of the Floquet states by using a global quantifier of the correlations. (These cat-state properties are strictly related to $\pi$-spectral pairing.) The Floquet states with the cat-state property are the correlated ones, and at the solvable point we find that are those with a finite local imbalance (for $s=1/2$). 
Moving away from the solvable point, we numerically show that the total amount of correlations for an average Floquet state still increases with system size, which indicates the presence of long-range correlations and further confirms the robustness of the time crystal phase.

In Appendix~\ref{swap:sec} we show how the spin flipping for the spin-1/2 case occurs, in Appendix~\ref{app:pidoub} we discuss the deep relation between $\pi$-spectral pairing and period doubling, {in Appendix~\ref{app:short-range-figures} we consider a different range of the interactions}, in Appendix~\ref{lisp:sec} we discuss in detail the imbalance distributions of the classical spin configurations for different values of $s$, {and in Appendix~\ref{corr_exact:app} we discuss the relation between local imbalance and correlations for $s=1/2$ at the solvable point}. In Sec.~\ref{sec:conclusions} we draw our conclusions.
%
\section{Model Hamiltonians}\label{sec:model}
\subsection{Spin-1/2 model}
\subsubsection{Description of the model}
We study a disordered spin-1/2 chain with power-law $ZZ$-interactions, nearest-neighbour $XX$-interactions and a longitudinal field which is subject to a periodic kick ($\hbar=1=T$). The periodic time-dependent Hamiltonian is
\begin{equation}\label{ht:eqn}
  \H(t) = \H_{\MBL} + \hK\sum_{n}\delta(t-n)\,,
\end{equation}
where the interaction Hamiltonian and the kicking Hamiltonian are respectively
\begin{align}\label{eq:Swap_Model_Full_Spin_Half}
    \H_{\MBL} &\equiv -\sum_{k=1}^{L-1} J(\hsigma_k^x\hsigma_{k+1}^x + \hsigma_k^y\hsigma_{k+1}^y) + \sum_{k=1}^{L-1}\sum_{q=k+1}^L V_{kq} \hsigma_k^z \hsigma_q^z +\nonumber\\
    &+\sum_{k=1}^L h_k^z\hsigma_k^z\,,\nonumber\\
    \hK &\equiv \left(\frac{\pi}{4}+\varepsilon\right) \sum_{k=1}^{L/2} \vsigma_{2k-1}\cdot\vsigma_{2k}\,,
\end{align}
where $\epsilon \ne 0$ describes imperfect swaps. So one can construct the time-evolution operator over one period {(the so-called Floquet operator)} as
\begin{equation}\label{foppa:eqn}
      \hU_F = e^{-\i\hK} e^{-\i\H_{\MBL}}\,.
\end{equation}
Here $\hsigma_k^{\beta=x,y,z}$ are the Pauli matrices; $V_{kq}$ follows a power-law behavior plus disorder,
\begin{equation}
V_{kq} = \frac{1}{\mathcal{N}_{L,\alpha}}\frac{\Tilde{V}_{kq}}{|k-q|^\alpha},
\end{equation}
where $\mathcal{N}_{L,\alpha}$ is a suitable Kac normalization constant~\cite{kac} which makes the energy extensive defined as
\begin{equation}
    \mathcal{N}_{L,\alpha} = 
    \begin{cases}
        1 & \text{if $\alpha\geq1$;}\\
        \ln L & \text{if $\alpha = 1$;}\\
        L^{1-\alpha} & \text{if $\alpha < 1$.}
    \end{cases}
\end{equation}
$\Tilde{V}_{kq}, h_k$ are independent random variables taken from the uniform box distributions $\Tilde{V}_{kq} \in [V/2,3V/2]$, $h_k \in [-h, h]$. {Power-law interactions are very important. Indeed nearest-neighbour interactions are not enough for time crystal, because they induce unwanted degeneracies between the Floquet states at the solvable point. These degeneracies spoil the time-translation symmetry breaking when one perturbs away from the solvable point, and are broken if one uses power-law interactions with disorder (see also Sec.~\ref{aminta:sec}).}

Let us focus on the kicking Hamiltonian $\hK$, with the associated unitary transformation corresponding to one kick $\hSw=e^{-\i\hK}$. The operator $\hSw$ can be factored into a product of unitary operators acting on neighboring sites (up to an irrelevant phase factor -- see Appendix~\ref{swap:sec})
\begin{equation}\label{swappero:eqn}
    e^{-\i\hK} = \prod_{k=1}^{L/2}\frac{1+\vsigma_{2k-1}\cdot\vsigma_{2k}}{2}\,.
\end{equation}
The operator $(1+\vsigma_i\cdot\vsigma_j)/2$ is the swap operator on two sites \cite{Zhang_2006}. This acts on a two-site product state by exchanging the states of the two sites (see Appendix~\ref{swap:sec} for a discussion)
\begin{equation}\label{eqn:swap}
    \frac{1+\vsigma_i\cdot\vsigma_j}{2}\ket{\psi_i,\psi_j} = \ket{\psi_j,\psi_i}\,.
\end{equation}
Therefore, the operator $\hSw$ exchanges the states of all odd sites with their following even sites. Since this operation is well defined only for pairs of spins, we restrict ourselves to the case of $L$ even. As we shall see, this protocol will show several differences in terms of both dynamics and spectrum with respect to the spin-flip protocol, such that it is not trivial to relate the two models. {Let us start considering a special point in the parameter space where the dynamics is exactly solvable.}

\subsubsection{Analysis of the solvable case ($J=\varepsilon=0$)}\label{solubo:sec}

\begin{figure}
    \centering
    \includegraphics[width=80mm]{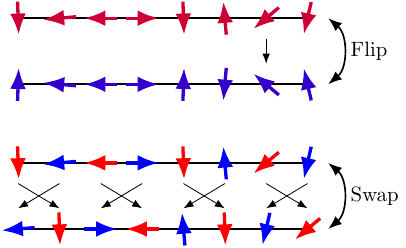}
        \caption{Comparison between swap and flip protocol. In the Flip case, all spins with non-zero magnetization are affected. In the swap case, all pairs of spins that are not aligned are affected.}
    \label{fig:Comparison_Flip_Swap}
\end{figure}

To understand the behavior of the model at the level of the dynamics and spectrum, it is instructive to look at the solvable case $J=\varepsilon=0$. In this case, $\H_{\MBL}$ commutes with all $\hsigma_k^z$ and $e^{-\i\hK}$ is a perfect swap. One can show that there is an \textit{exact} period-doubling dynamics, which can be seen at the level of the operators themselves
\begin{equation}
    \hsigma_k^z(T) = \hU_F^{\dagger}\hsigma_k^z \hU_F = \hsigma_{\overline{k}}^z,
\end{equation}
where $\overline{k} = k+1$ if $k$ is odd and $\overline{k} = k-1$ is $k$ is even. Therefore, the magnetization serves as {an} order parameter, and period-doubling will be observed for all initial states where, for some $k$, $\braket{\hsigma_k^z} \neq \braket{\hsigma_{\overline{k}}^z}$, since the local magnetization will oscillate between its initial value and that of its paired neighbour. {We graphically show this behavior in Fig.~\ref{fig:Comparison_Flip_Swap} and compare it with the more known case of the spin flipping.}

We note that the period-doubling will be a measurable effect in the thermodynamic limit only if a finite fraction of sites satisfies the above condition, i.\,e., there is an extensive quantity of not aligned spins.

As an example, one can imagine an initial state where the odd spins lie in the $xy$ plane -- such that $\braket{\hsigma_{2k-1}^z(t=0)} = 0$ -- and the even spins lie in the $z$ direction -- and so $\braket{\hsigma_{2k}^z(t=0)} = \pm 1$. For the odd ones, the interaction part results in a dephasing of the spins in the $xy$-plane~\cite{PhysRevB.90.174302} and the $z$-magnetization remains zero. For the even ones, being $J=0$, the interaction part of the Hamiltonian does not change the spins pointing along the $z$ direction. So, due to the periodic swap, the magnetization at each site will exhibit period-doubling by oscillating between $0$ and $\pm 1$.

Furthermore, certain configurations may display period-doubling restricted only to a part of the chain, such as one which is part anti-ferromagnetic ($\{+-,+-,\cdots\}$) and part ferromagnetic ($\{++,++,\cdots\}$). In this case the order parameter $\braket{\hsigma_k^z(t)}$ oscillates only in a certain part of the chain. Nevertheless, if the anti-ferromagnetic part is extensive, then we still consider such a state as breaking the time-translation symmetry of the system.

We highlight that the number of spin configurations where no period-doubling takes places increases with system size as $2^{L/2}$ and the number of remaining configurations is $2^L - 2^{L/2}$. As a consequence, the configurations which display no period-doubling throughout the entire chain must satisfy a very special condition, as their fraction vanishes with increasing system size as $2^{-L/2}$ over the total number of spin configurations.

Thus, we find that this swapping dynamics is in general quite different than the prototypical model with spin-flip~\cite{Nayak2016}, where the magnetization always (and only) changes sign, independently of neighbouring spins (see Fig.~\ref{fig:Comparison_Flip_Swap}).
\subsection{Spin-1 model and beyond}\label{sec:Spin-1_Model}

The interesting aspect of spin swapping is that one can generalize the time-crystal behavior of the spin-1/2 model to generic local spin magnitude. The price to pay is that, for increasing spin magnitude $s$, the spin-swapping operator becomes more and more complicated~\cite{Zhang_2006}. 

Let us focus here on the spin-1 case ($s=1$).
The Hamiltonian for this case has the same structure as Eq.~\eqref{ht:eqn}. The interaction part has much the same form {as} the spin-1/2 case, although the periodic swap also involves higher powers of the spin operators
\begin{equation}\label{eq:Swap_Model_Full_Spin_One}
    \begin{split}
        \H_{\MBL} &= -\sum_{k=1}^{L-1} J(\hS_k^x\hS_{k+1}^x + \hS_k^y\hS_{k+1}^y) \\ 
        &\quad {}+ \sum_{k=1}^{L-1}\sum_{q=k+1}^L \hS_k^z \hS_q^z +\sum_{k=1}^L h_k^z\hS_k^z,\\
        \hK &= \left(\frac{\pi}{2}+\varepsilon\right) \sum_{k=1}^{L/2} \left[(\vS_{2k-1}\cdot\vS_{2k})^2 + \vS_{2k-1}\cdot\vS_{2k}\right],\\
        \hU_F &=  e^{-\i\hK}e^{-\i\H_{\MBL}}.
    \end{split}
\end{equation}
and the operators $\hS_k^{\beta=x,y,z}$ are now spin-1 operators.

$V_{kq},h_k$ have the same distributions as in the spin-1/2 case. Of course, even though the Hamiltonian is similar in form, the spin-1 chain has a different behavior with respect to the spin-1/2, due to the additional  state with $s_k = 0$~\cite{PhysRevB.100.144423}.

The main difference with the spin-1/2 case is that the number of classical configurations providing period doubling is given by $3^L-3^{L/2}$ [for generic local spin magnitude $s$ it is $(2s+1)^L-(2s+1)^{L/2}$]. So, the ratio of configurations that provide no period doubling is for generic spin $(2s+1)^{-L/2}$. It is smaller than in the spin-1/2 case, and it decreases for increasing value of the local spin $s$. Furthermore, for a random spin configuration there will be a larger fraction of sites which participate to the period-doubling (see see Appendix~\ref{lisp:sec}).

It is important to define the spin basis $\ket{\{s_k\}}$. In the spin-1/2 case this basis is such that $\hsigma_k^z\ket{\{s_k\}} = s_k\ket{\{s_k\}},\;s_k \in\{ +,-\}$, while in the generic case $\hS_k^z\ket{\{s_k\}} = s_k\ket{\{s_k\}},\;s_k \in\{ -s,-s+1,\ldots,s-1,s\}$. Another important property of our models is that the {dynamics} preserves the total $z$ magnetization ($\hat{\mathcal{S}}_z = \frac{1}{2}\sum_{k=1}^L\hsigma_k^z$ for the spin-1/2 case, $\hat{\mathcal{S}}_z = \sum_{k=1}^L \hS_k^z$ otherwise), which will allow us to restrict our numerics to a subspace with fixed $\mathcal{S}_z$. In the following are we going to study analytically and numerically different aspects of the period-doubling behavior.
%
\section{Local imbalance}\label{li:sec}
In order to easily identify which initial states lead to period-doubling we define the following operator, which we call Local Imbalance ($\hLI$). Our aim is to find an operator that, on classical spin configurations, counts how many pairs of swapped spins have different values of the magnetization $\braket{\{s_k\}|\hLI|\{s_k\}} = \frac{1}{L/2}\sum_k (1 - \delta_{s_{2k-1},s_{2k}})$. In the spin-1/2 case, this operator is
\begin{equation}\label{eq:Def_Local_Imbalance}
    \hLI = \frac{1}{L/2}\sum_{k=1}^{L/2} \frac{(\hsigma_{2k}^z - \hsigma_{2k-1}^z)^2}{4}\,,
\end{equation}
while, in the spin-1 case, it is
\begin{align}\label{eq:Def_Local_Imbalance_Spin-1}
    \hLI &= \frac{1}{L/2}\sum_{k=1}^{L/2}\big[ (\hS_{2k-1}^z)^2 + (\hS_{2k}^z)^2 \nonumber\\
        &- \frac{1}{2}\hS_{2k-1}^z\hS_{2k}^z(1+3\hS_{2k-1}^z\hS_{2k}^z) \big]\,.
\end{align}
{This quantity -- inspired by the charge-imbalance in fermionic chains~\cite{exp_mbl1,Kohlert_2019} -- measures the difference in alignment (spin analogue of the fermionic occupation) in all the $L/2$ even-odd pairs, regardless of their sign (hence the square).}

{Let us first focus on the solvable case, and consider the behavior of $\hLI$ on the spin basis $\ket{\{s_k\}}$. When $\braket{\{s_k\}|\hLI|\{s_k\}}=1$, pairs have different spin values ($s_{2k} \neq s_{2k-1}$), the swap makes spins oscillate between two values, and the dynamics is the same as for the spin-flip protocol. When $\braket{\{s_k\}|\hLI|\{s_k\}}=0$, all pairs have the same spin value, and the state is left unchanged by the periodic swap. In the general case $0<\braket{\{s_k\}|\hLI|\{s_k\}}<1$ only certain pairs will give rise to period-doubling. This is in contrast with the spin-flip case, where any classical configuration gives rise to period-doubling oscillations at the solvable point.}

Much of this also holds for states \textit{close} to the spin basis. For instance, focus on the spin-1/2 case and consider the product state where spins have alignment close to the $z$ direction, {\em i.\,e.}, $\bigotimes_{k=1}^{L} (\cos(\theta_k)\ket{+} + \sin(\theta_k)\ket{-})$ for $\theta_k$ equal to $0$ or $\pi$ up to small corrections  $\delta\theta\ll 1$. This state clearly displays period-doubling in every site and at leading order in $\delta \theta$ we have $\LI = 1 - C(\delta\theta)^2$ with $C$ a constant. Therefore, in this case, the high value of local imbalance accurately predicts the presence of period-doubling. 


Notice that at the solvable point $\hat{\LI}$ commutes with $\hat{U}_F$, meaning that it provides also a good quantum number for the eigenstates of $\hat{U}_F$ and is a useful signature of their properties, as we shall see below. 
It is also important to remark that, for increasing local spin magnitude $s$, the average local imbalance of the states in the spin basis increases, and the variance decreases, as we show in detail in Appendix~\ref{lisp:sec}. This is interesting because initializing the system with a spin state with nonvanishing imbalance leads to period-doubling oscillations lasting forever, at least for the solvable case. 

In the next section we move away from the solvable point and show that there are still period-doubling oscillations that last for a time increasing with the system size, so we have not a special isolated point but a robust time-crystal phase.
\section{Period-doubling oscillations} \label{peridolo:sec}
A time-crystal behavior is signaled by persistent period-doubling oscillations of some observables. In order to speak about a phase and not an isolated point, these oscillations must be robust to changes of the parameters. Let us consider what happens when we move away from the solvable point and take  $J\neq 0$. We are going to show that there is an interval of $J$ where period-doubling oscillations appear, and last a time that exponentially increases with the system size, for many possible initial states. We see that both in the case of spin-1/2 model (Sec.~\ref{spin12_dyn:sec}), and spin-1 model (Sec.~\ref{spin1dyn:sec}). 

In the current and following sections we shall consider the disorder average of various quantities, defined as $\overline{(\ldots)} = \frac{1}{N_d}\sum_{i=1}^{N_d} (\ldots)$. The chosen number of disorder realizations $N_d$ will typically get smaller with $L$, since we take advantage of the self-averaging behaviour - specifically, we take $N_d=20480 \cdot 2^{1-L/2}$, unless otherwise mentioned, which allows for sufficiently small statistical errors. 

Furthermore, we fix $V = 3$ and $h_z = 16$ in all our computations and mostly focus on the perfect swap case $\varepsilon=0$ - except in certain cases, which we explicitly mention. We shall consider two choices of interaction range: $\alpha=0.5$ (long-range) and $\alpha=3$ (short-range). Since we have found that the two values typically have only very slight quantitative differences, we will mostly focus on the $\alpha=0.5$ case, and leave the $\alpha=3$ case in Appendix~\ref{app:short-range-figures}.

In the entire section we will focus on stroboscopic dynamics, i.\,e., the time $t=nT$ will be always an integer multiple ($n\in\mathbb{N}$) of the period $T=1$.
\subsection{Spin-1/2 case}\label{spin12_dyn:sec}
\begin{figure}
    \centering
    \includegraphics[width=80mm]{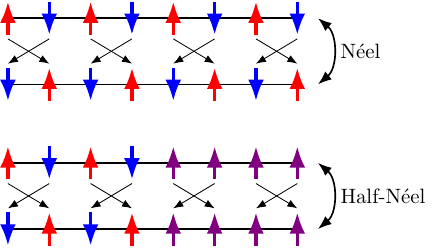}
        \caption{Initial Néel (upper panel) and Half-Néel (lower panel), with their respective dynamics, which are exact only in the solvable case.}
    \label{fig:Neel_Dynamics_Picture}
\end{figure}

In this section we consider the problem of period-doubling dynamics by looking at two initial states in the spin configuration with finite $\LI$, such that $\braket{\hsigma_k^z} \neq \braket{\hsigma_{\overline{k}}^z}$ over a finite fraction of the total sites. Specifically, we consider an initial Néel state $\ket{+-+-\cdots}$ (fully anti-ferromagnetic) and a “half-Néel” state $\ket{+-+-\cdots++++\cdots}$ (anti-ferromagnetic on the length $l_{\AF} = 2\lfloor L/4\rfloor$, and the rest ferromagnetic), as pictured in Fig.~\ref{fig:Neel_Dynamics_Picture}.

With the first state one gets the same kind of dynamics as the one observed in the spin-flip protocol, since all spins get inverted after each kick. Indeed, it has maximal local imbalance $\LI=1$. With the second initial state, “half-Néel”, instead one gets period-doubling dynamics only in the first half of the chain and $\LI=\lfloor L/4\rfloor /(L/2)$.

In order to monitor the period-doubling dynamics obtained with these initial states, we consider the following quantity, analogous to the one used in~\cite{Nayak2016,note_theo}
\begin{equation}\label{eq:Z_Dynamics_Definition}
    \begin{split}
        Z(t) = \frac{1}{L/2}\sum_{k=1}^{L/2} \sign(\braket{\hsigma_{2k-1}^z(0)} - \braket{\hsigma_{2k}^z(0)}) \times\\
        \times (\braket{\hsigma_{2k-1}^z(t)} - \braket{\hsigma_{2k}^z(t)})\,.
    \end{split}
\end{equation}
$Z(t)$ is defined in such a way that it is always non-negative at $t=0$ and at the solvable point ($J=\varepsilon=0$) oscillates between a positive and negative value every period $Z(t) = (-1)^t Z(0)$, displaying in this way period doubling. It has zero contributions from regions where $\braket{\hsigma_{2k-1}^z} = \braket{\hsigma_{2k}^z}$ {and due to the normalization factor $L/2$, only states with an extensive quantity of unaligned spin pairs have a finite $Z(0)$ in the thermodynamic limit.}

\begin{figure}
    \centering
    \begin{overpic}[width=80mm]{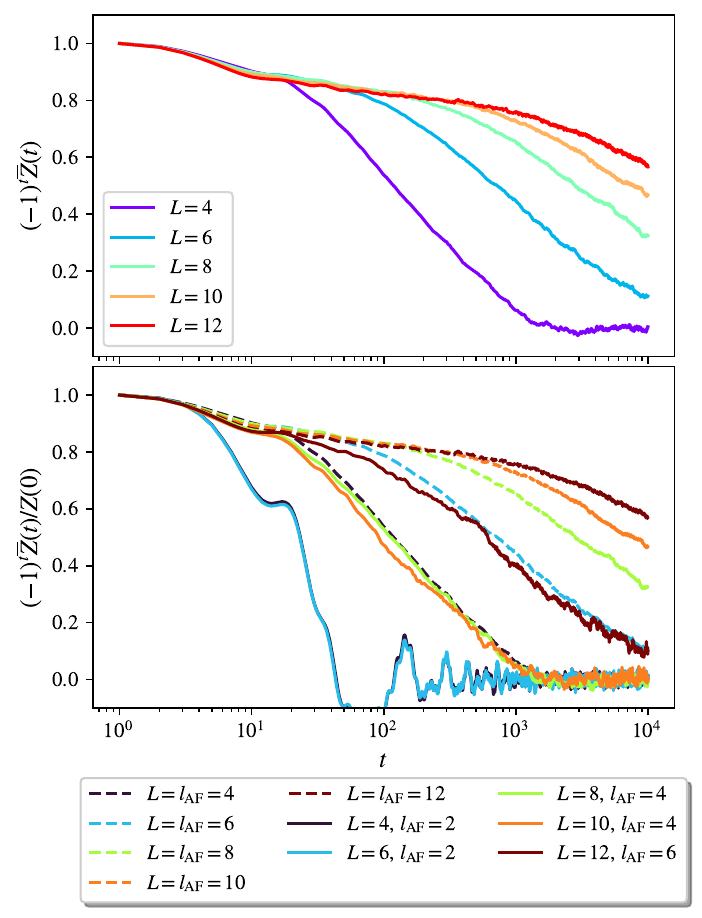}\put(0,95){(a)}\put(0,55){(b)}\end{overpic}
    \caption{Spin-1/2 model. (Panel a) Dynamics of $(-1)^t\overline{Z}(t)$ for initial Néel state and different system sizes. (Panel b) Dynamics of $(-1)^t\overline{Z}(t)$ for initial Half-Néel state (solid line) and Néel state (dashed line) and different system sizes. Initial values have all been normalized to $+1$ to make the comparison clearer. Numerical parameters: $J=0.1$, $\varepsilon = 0.01$.}
    \label{fig:SpinHalf_Z_dynamics}
\end{figure}

Let us average over disorder and focus on the quantity $(-1)^t\overline{Z}(t)$, that in presence of period doubling is constantly positive and does not change sign at every period. In Fig.~\ref{fig:SpinHalf_Z_dynamics}(a) we show some examples of $(-1)^t \overline{Z}(t)$ versus $t$, for different system system sizes and initializing with the Néel state. We can qualitatively see that $(-1)^t\overline{Z}(t)$ decays over a time scale that increases with system size. 

\begin{figure}
    \centering
    \includegraphics[width=\linewidth]{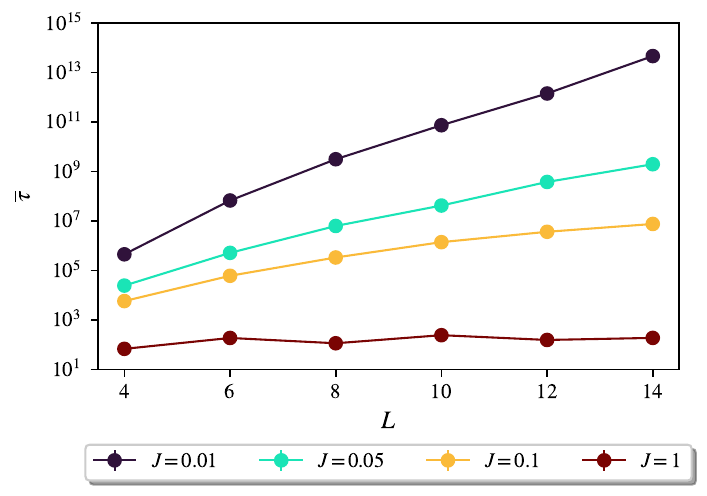}\\
    \caption{Spin-1/2 model. Decay times $\overline{\tau}$ of $(-1)^tZ(t)$ for an initial Néel state. Here we have considered time evolutions while skipping over $n\geq 1$ periods, such that the time step stays below $\sim 1\%$ of the standard deviation of $\tau$.}
    \label{fig:Decay_Times_Neel_alpha0.5}
\end{figure}

{Let us study this behavior of the decay time more quantitatively.} For a given disorder realization, we define the decay time $\tau$ as the first time where $Z(t)$ no longer manifests period-doubling behaviour, i.\,e., $Z(\tau-1)Z(\tau) > 0$ (if multiple periods are skipped the condition becomes $(-1)^n Z(\tau-n)Z(\tau) < 0$). {In Fig.~\ref{fig:Decay_Times_Neel_alpha0.5} we show $\overline{\tau}$ versus $L$ for different system sizes. We notice that the decay time exponentially increases with system size for a range of parameters. This is a signature of stable period-doubling in the thermodynamic limit \cite{PhysRevResearch.2.012003,Potter2017}, meaning that there is time crystal behaviour in this range of parameters.}

We show $\overline{Z}(t)$ versus $t$ initializing with the half-Néel state in Fig.~\ref{fig:SpinHalf_Z_dynamics}(b). We notice that the curves do not seem to depend on the total length of the chain $L$, but rather on the length of the anti-ferromagnetic part $l_{\AF}=2\lfloor L/4\rfloor$. Each curve for the N\'eel state closely follows the one for the half Néel state if the antiferromagnetic parts are of the same size.

\begin{figure}
    \centering
    \includegraphics[width=\linewidth]{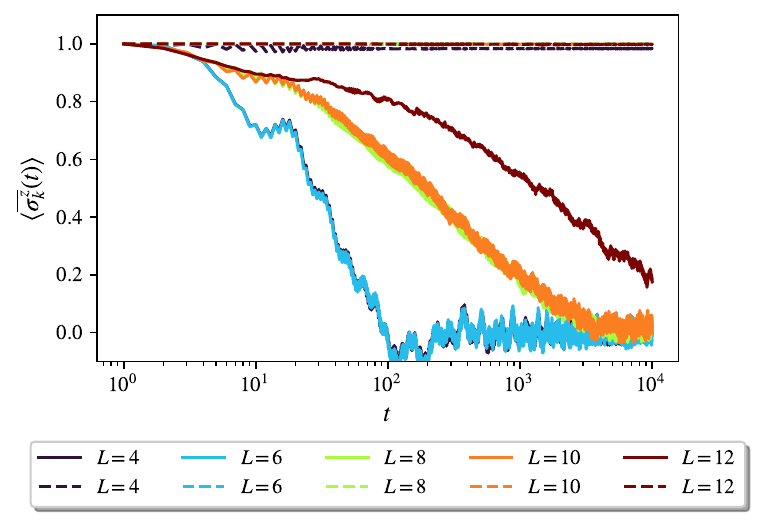}
    \caption{Spin-1/2 model. Comparison of $\braket{\overline{\hsigma_k^z}(t)}$ at $k=L-\lfloor L/4\rfloor$ (dashed line) and $(-1)^t\braket{\overline{\hsigma_k^z}(t)}$ at $k=\lfloor L/4\rfloor$ (solid line), up to $t=10^4$, for an initial Half-Néel state. Parameters: $J=0.1$}
    \label{fig:Half_Néel_Comparison_sigmaz_dynamics}
\end{figure}

It is also worth noticing that even though $Z(t)$ decays to zero at long times, the local magnetizations $\braket{\hsigma_k^z(t)}$ themselves do not: The ones in the ferromagnetic part are frozen (see Fig.~\ref{fig:Half_Néel_Comparison_sigmaz_dynamics}). This is an effect of many-body localization (see Sec.~\ref{regdyn:sec}) that hinders energy absorption and infinite-temperature thermalization and induces space localization of excitations. 

\subsection{Spin-1 case} \label{spin1dyn:sec}

\begin{figure}
    \centering
    \begin{overpic}[width=80mm]{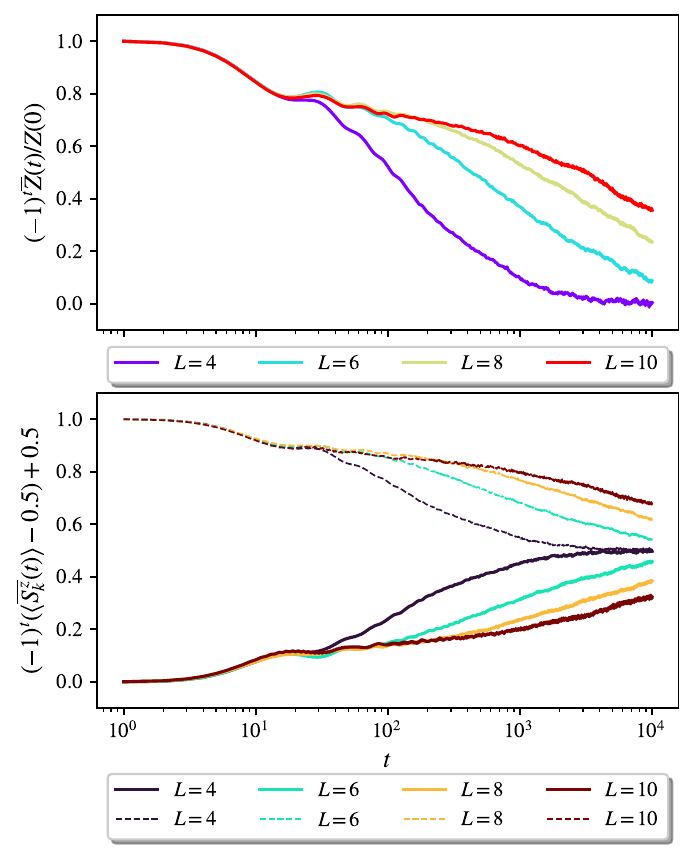}\put(0,95){(a)}\put(0,55){(b)}\end{overpic}
    \caption{Spin-1 model. (Panel a) Dynamics of $(-1)^t\overline{Z}(t)$ for an initial Up-Zero state. (Lower panel) Dynamics of $(-1)^t(\braket{\overline{\hS_k^z(t)}}-0.5)+0.5$ at $k=2\lfloor L/4\rfloor$ (solid line) and $k=2\lfloor L/4\rfloor+1$ (dashed line) for an initial Up-Zero state. Parameters: $J=0.1$.}
    \label{fig:UpZero_dynamics}
\end{figure}

Let us now look at the dynamics of the spin-1 chain, where once again we require that in the initial state $\braket{\hS_k^z} \neq \braket{\hS_{\overline{k}}^z}$ for an extensive number of sites. Specifically, we consider the initial state “Up-Zero” given by the alternating sequence $\ket{+1,0,+1,0,\cdots}$. This state gives an example of a “non-flip” dynamics already within the spin configuration basis, in the sense that each site oscillates between $\braket{\hS_k^z(t)} = 0$ and $\braket{\hS_k^z(t)} = +1$ (solvable case) and does not reduce to a spin-flip. Furthermore, since the local magnetization changes value at each site, its local imbalance [Eq.~\eqref{eq:Def_Local_Imbalance_Spin-1}] has maximal value $1$.
Once again, in order to study the dynamics of the system we use the observable $Z(t)$ defined in Eq.~\eqref{eq:Z_Dynamics_Definition}, with $\hS_k^z$ instead of $\hsigma_k^z$. The behavior of its disorder average $\overline{Z}(t)$ versus $t$ for different system sizes is shown in Fig.~\ref{fig:UpZero_dynamics}(a).

\begin{figure}
    \centering
    \includegraphics[width=\linewidth]{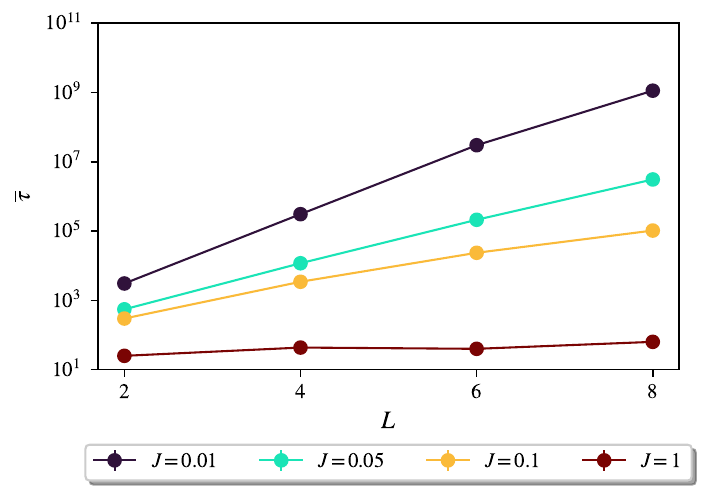}\\
    \caption{Spin-1 model. Decay times $\overline{\tau}$ of $(-1)^tZ(t)$, averaged over disorder for an initial Up-Zero state.}
    \label{fig:Decay_Times_UpZero_alpha0.5}
\end{figure}

We find again a range of parameters where $\overline{Z}(t)$ oscillates between a positive and negative value and decays to zero in a time exponentially increasing with the system size, as we can see in the behavior of  $\overline{\tau}$ versus $L$ shown in Fig.~\ref{fig:Decay_Times_UpZero_alpha0.5}. Again, this means that there is time crystal behaviour in this range of parameters.

The period-doubling oscillations and the fact that they last for a time exponential in the system size can be seen also at the level of the expectations of the onsite magnetizations [see Fig.~\ref{fig:UpZero_dynamics}(b)]. Specifically, by taking the middle of the chain as representative, we see that both sites (starting at either $\braket{S_k^z} = 0$ or $1$) oscillate around $\braket{S_k^z} = 0.5$ and decay towards this middle value at exponentially long times. 


\section{Properties of the Floquet spectrum}\label{floq:sec}
In order to understand the behavior of the model throughout the entire Hilbert space, it is useful to look at the Floquet spectrum. Thus, we discuss the properties of the quasienergies $\{\mu_\beta\}$ (Sec.~\ref{regdyn:sec},~\ref{aminta:sec}) and the Floquet states $\{\ket{\psi_\beta}\}$ (Sec.~\ref{amanta:sec}) in relation with thermalization properties and period doubling.

\subsection{Regular dynamics}\label{regdyn:sec}

In order to see persistent period doubling, there must be no thermalization of local observables. (Local thermalization is a common behavior of many-body quantum systems~\cite{polkovnikov83nonequilibrium}). The dynamics of the system must be regular, that's to say similar to an integrable quantum system. Here this effect can be provided by disorder, that is known to break ergodicity and induce many-body localization~\cite{Pal_PhysRevB10,RevModPhys.91.021001,PhysRevB.97.214205}. 

In order to check the range of $J$ where integrable-like behavior occurs, we {diagonalize the Floquet operator defined in Eq.~\eqref{foppa:eqn}} as $\hU_F\ket{\psi_\beta} = e^{-\i\mu_\beta}\ket{\psi_\beta}$. The eigenvectors $\ket{\psi_\beta}$ are the so-called Floquet states and the phases of the eigenvalues $\mu_\beta$ are the corresponding quasienergies~\cite{shirley1965solution,Sambe1973}. Let us evaluate the average level spacing ratio $\braket{r}$~\cite{Pal_PhysRevB10} defined as
\begin{equation}\label{r:eqn}
  \braket{r} = \frac{1}{\mathcal{N}}\sum_{\beta=1}^{\mathcal{N}_M-2}\overline{\left[\frac{\min(\mu_{\beta+2}-\mu_{\beta+1},\mu_{\beta+1}-\mu_\beta)}{\max(\mu_{\beta+2}-\mu_{\beta+1},\mu_{\beta+1}-\mu_\beta)}\right]}
\end{equation}
where quasienergies are restricted to the first Floquet Brillouin zone~\cite{Russomanno_PRL12,rudner_2012} $[-\pi,\pi]$ (they are periodic of period $2\pi$) and taken in increasing order. $\mathcal{N}$ is the dimension of the sector of the Hilbert space with vanishing total magnetization, $s_z = 0$. We consider in Eq.~\eqref{r:eqn} only the quasienergies $\mu_\beta$ corresponding to Floquet states in this sector. In this way we are restricting to an irreducible invariant subspace of the Hamiltonian, a condition required in order for the  distribution of the level spacings $\delta_\beta = \mu_{\beta+1}-\mu_\beta$ (and the related ratio $\braket{r}$) to be a meaningful ergodicity indicator~\cite{Berry_LH84}. 

When the driven system is ergodic, i.\,e., locally thermalizing with infinite temperature~\cite{Lazarides_PRE14}, the Floquet operator $\hU_F$ belongs to the circular orthogonal ensemble (COE) of symmetric unitary matrices (because of the time-reversal invariance)~\cite{Rigol_PRX14,Haake:book,eynard2018random,PhysRevLett.110.084101}. In this case, the distribution of the level spacings $\delta_\beta =\mu_{\beta+1}-\mu_\beta$ is of the COE type and the average level spacing ratio acquires the value $\braket{r}_{\rm COE}\simeq 0.5269$. 

A level spacing distribution of the Poisson type, instead,  corresponds to an integrable dynamics~\cite{Berry_PRS76} and is observed for instance for many-body-localized systems, both autonomous~\cite{Pal_PhysRevB10} and periodically-driven~\cite{ponte2015many}. Like the present ones, these are systems with disorder, and they never thermalize due to the existence of a superextensive number of localized integrals of motion~\cite{PhysRevLett.111.127201,ros2015integrals,Imbrie2016} that deeply affect dynamics, and forbid thermalization of local observables. In the periodically-driven case this hinders energy absorption and avoids infinite-temperature thermalization. Integrability and Poisson level-spacing distribution correspond to an average level spacing ratio $\braket{r}_P\simeq 0.386$, and due to the presence of disorder we expect to see them here in some range of parameters.

\begin{figure}
    \centering
    \begin{tabular}{c}
         \begin{overpic}[width=80mm]{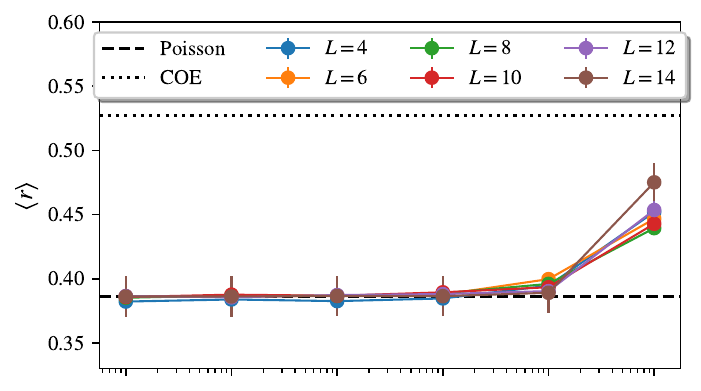}\put(0,48){(a)}\end{overpic}\vspace{-0.3cm}\\
         \begin{overpic}[width=80mm]{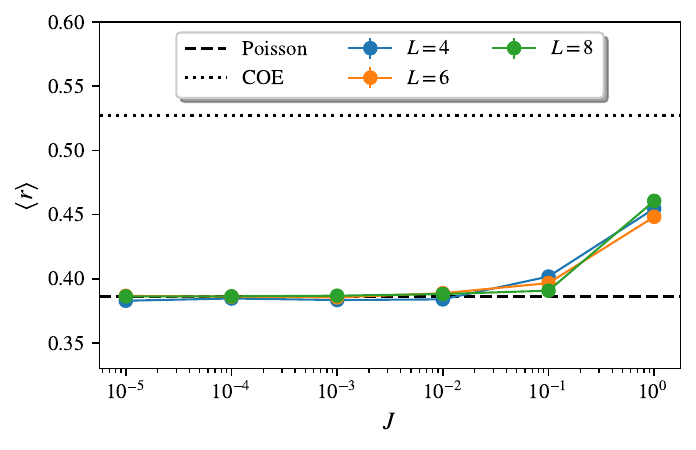}\put(0,55){(b)}\end{overpic}\vspace{-0.5cm}
    \end{tabular}
    \caption{Average level spacing ratio $\braket{r}$ versus $J$ in the spin-1/2 case (panel a) and spin-1 case (panel b) for different $L$ and $\alpha=0.5$.}
    \label{fig:Gap_Ratio}
\end{figure}

To inquire this point, we plot $\braket{r}$ versus $J$ {for different system sizes in the spin-1/2 model [Fig.~\ref{fig:Gap_Ratio}(a)] and the spin-1 model [Fig.~\ref{fig:Gap_Ratio}(b)].} When the size $L$ is large enough, we see that there is an interval where $\braket{r}$ is stuck to the Poisson value. So there is a parameter region -- in all the considered cases $J\lesssim 10^{-1}$ -- where the dynamics is regular. This property persists as the system size is increased (at least for the system sizes we can numerically attain). {Then we can talk about an integrable-like phase where there is no thermalization and} a nontrivial dynamics like the one leading to persistent period-doubling oscillations can appear.

%

\subsection{$\pi$-spectral pairing of the quasienergies}\label{aminta:sec}
\subsubsection{Theoretical discussion}
A necessary condition for the presence of period doubling is the so called $\pi$-spectral pairing~\cite{Nayak2016,PhysRevLett.100.070502,jian2015longrange,Keyserlingk2016,Khemani2016}, i.\,e., each quasienergy must have a partner shifted by $\pi$. (See~\cite{Surace_2019} and Appendix~\ref{app:pidoub} for an analysis of how $\pi$-spectral pairing gives rise to period doubling). At the solvable point our spin-1/2 model shows $\pi$-spectral pairing, as can be easily checked. 

Let us define $\ket{\{s_{\overline{k}}\}}$ as the transformed configuration of $\ket{\{s_k\}}$ under spin swap, and consider that the energy of the classical configuration is
$$
  E(\{s_{k}\}) = \sum_{k=1}^{L-1}\sum_{q=k+1}^L V_{kq} s_k^z s_q^z+\sum_{k=1}^L h_k^z s_k^z\,.
$$
We can write the action of the Floquet operator Eq.~\eqref{foppa:eqn} on the spin configurations mentioned above
\begin{align}
  &\hU_F\ket{\{s_k\}} = \nep^{-\i E(\{s_k\})} \ket{\{s_{\overline{k}}\}}\,,\nonumber\\
  &\hU_F\ket{\{s_{\overline{k}}\}} = \nep^{-\i E(\{s_{\overline{k}}\})} \ket{\{s_k\}}\,.
\end{align}
 From these relations {one can diagonalize $\hU_F$} and find the Floquet states as  
\begin{equation}  \label{cat:eqn}
  \ket{\psi_\beta^\pm} = \frac{1}{\sqrt{2}}[ e^{\i E(\{s_k\})/2}\ket{\{s_k\}} \pm e^{\i E(\{s_{\overline{k}}\})/2}\ket{\{s_{\overline{k}}\}}]\,,
\end{equation}
and the corresponding quasienergies respectively as
\begin{align}
  &\mu_\beta^+ = [E(\{s_k\}) + E(\{s_{\overline{k}}\})]/2\,,\nonumber\\
  &\mu_\beta^- = \mu_\beta^+ + \pi\,,\quad {\rm whenever} \quad\{s_{\overline{k}}\}\neq \{s_{k}\}\,.
\end{align}
 Therefore a part of the Floquet states show $\pi$-spectral pairing, {\em i.\,e.}, $\mu_\beta^--\mu_\beta^+=\pi$. This is valid only for the configurations such that $\{s_{\overline{k}}\}\neq \{s_{k}\}$, that are $(2s+1)^L-(2s+1)^{L/2}$ in number (see Sec.~\ref{solubo:sec}), that is, the overwhelming majority. {If the configurations $\{s_k\}$ and $\{s_{\overline{k}}\}$ differ over an extensive number of sites, the Floquet states $\ket{\psi_\beta^\pm}$ are cat states and are globally different from each other.}

We can interpret the oscillations of the magnetization as Rabi oscillations of a spin configuration $\ket{\{s_k\}}$ with its swapped partner $\ket{\{s_{\overline{k}}\}}$. Indeed these configurations are given by linear combinations of the two Floquet states $\ket{\psi_{\beta,+}} \pm \ket{\psi_{\beta,-}}$, and the relative phase $\pm 1 = \mp e^{\i\pi}$ gets inverted every time the Floquet operator acts on the state. Specifically, if all eigenstates are $\pi$-paired, this guarantees that there will be period-doubled oscillations for any generic initial state.

We note here that the presence of disorder and power-law interactions is fundamental, as it removes all degeneracies between distinct quasienergies at the solvable point. Due to this gap, the application of a small perturbation is expected to preserve the structure of the Floquet states and the spectral pairing. 

More precisely, states that {differ only locally ({\em i.\,e} are coupled {\em only} by operators localized on a specific small range of sites~\cite{Nayak2016})} 
will have different quasienergies. {This property is called local spectral gap, and in the presence of this property} we expect that for all eigenstates \textit{local perturbations perturb locally}, in a way that is analogous to the ground states of gapped systems \cite{Bachmann2012-bd}: {These perturbations couple globally different Floquet states at order $\mathcal{O}(L)$ in perturbation theory, and so  do not mix them at any perturbative order in the limit $L\to\infty$~\cite{tesi_fede,hastings2010quasiadiabatic}.} This implies that there exists a local unitary transformation which relates the original eigenstates to the perturbed eigenstates~\cite{Bauer_2013,ponte2015many} and in particular the properties of the Floquet spectrum are conserved under local perturbations~\cite{Nayak2016,tesi_fede}. As a consequence, we expect that even in the presence of perturbations ($J,\varepsilon\neq 0$) there will still be $\pi$-spectral pairing and thus a time crystal phase over a finite region of parameters. This expectation is fully confirmed by the numerical results we are going to discuss.

\subsubsection{Numerical analysis}\label{numerello:sec}
\begin{figure}
    \centering
    \includegraphics[width=50mm]{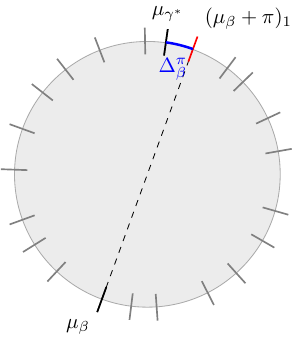}
    \caption{Choice of $\pi$-paired eigenvalue $\mu_{\gamma^*}$ corresponding to a given $\mu_\beta$ and its $\pi$-shifted spectral gap $\Delta_\beta^\pi$.}
    \label{fig:pi-Spectral_Pairing}
\end{figure}

To study the persistence of the spectral pairing outside the solvable case, we consider the following quantities~\cite{PhysRevB.94.085112,PhysRevB.95.214307}
\begin{equation}\label{eq:pi-Spectral_Pairing}
    \begin{split}
        \Delta_\beta^0 &= \mu_{\beta+1} - \mu_{\beta}\,;\\
        \Delta_\beta^\pi &= \min_\gamma \abs{\mu_\gamma - (\mu_{\beta}+\pi)_1}\,. 
    \end{split}
\end{equation}
All quasienergies are taken in the first Brillouin zone $\mu_\beta,(\mu_\beta+\pi)_1 \in (-\pi,\pi)$ and the differences between quasi-energies are also taken $\text{mod  }2\pi$, i.\,e., we evaluate $\Delta_\beta^\pi$ by minimizing the $\pi$-distance between quasienergies on the unit circle, as shown in Fig.~\ref{fig:pi-Spectral_Pairing}.

\begin{figure}
    \centering
    \begin{tabular}{c}
         \begin{overpic}[width=80mm]{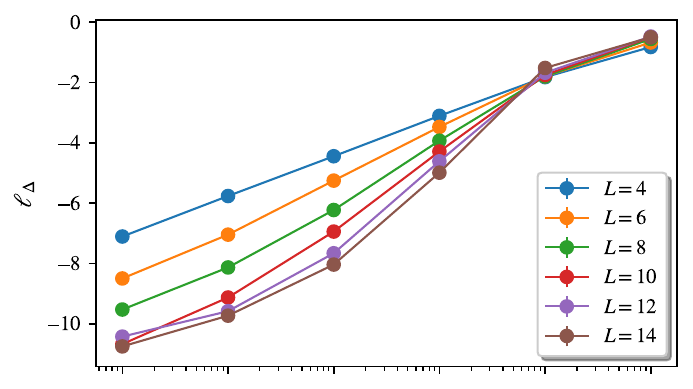}\put(0,48){(a)}\end{overpic}\vspace{-0.3cm}\\
         \begin{overpic}[width=80mm]{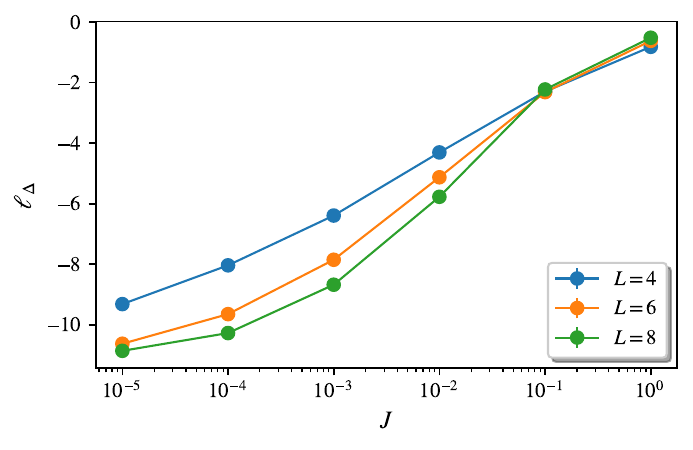}\put(0,50){(b)}\end{overpic} \vspace{-0.5cm}
    \end{tabular}
    \caption{Spectral-paring parameter $\ell_\Delta$ versus $J$ in the spin-1/2 case (panel a) and spin-1 case (panel b) for different $L$ and $\alpha=0.5$.}
    \label{fig:Spectral_Pairing_Difference_Log_Delta}
\end{figure}

In case the system features $\pi$-spectral pairing in the thermodynamic limit, the $\pi$-shifted spectral gaps $\Delta_\beta^\pi$ should scale to zero with system size faster than the spectral gaps $\Delta_\beta^0$. In order to quantitatively probe this property, we numerically diagonalize $\hU_F$ in the $\mathcal{S}_z=0$ subspace and compute the logarithmic averages~\cite{PhysRevB.94.085112,note:log} 
\begin{align}
	&\braket{\log_{10}\Delta^\pi}\equiv\frac{1}{\mathcal{N}}\sum_\beta \overline{\log_{10} \Delta_\beta^\pi}\,,\quad{\rm and}\nonumber\\
	&\braket{\log_{10}\Delta^0}\equiv \frac{1}{\mathcal{N}}\sum_\beta \overline{\log_{10} \Delta_\beta^0}\,,
\end{align}
where the sum over $\beta$ is restricted to the quasienergies of Floquet states in the sector of the Hilbert space with $\mathcal{S}_z = 0$.
%
In order to understand if there is $\pi$ spectral pairing in the thermodynamic limit, we consider the {spectral-pairing parameter} 
\begin{equation}\label{speppe:eqn}
	\ell_\Delta \equiv \braket{\log\Delta^\pi} - \braket{\log\Delta^0}\,.
\end{equation}
In the case of spectral pairing this difference should decrease with system size $L$. We plot $\ell$ versus $J$ for different values of $L$ in Fig.~\ref{fig:Spectral_Pairing_Difference_Log_Delta}(a) for the spin-1/2 and Fig.~\ref{fig:Spectral_Pairing_Difference_Log_Delta}(b) for the spin-1 case.

We see that in all the considered cases there is a threshold value for $J\sim 10^{-1}$ where all the curves appear to cross.  Below this threshold, $\ell_\Delta$ decreases with the system size, marking the presence of $\pi$-spectral pairing (the results at large $L$ and very small $J$ are flawed by the finite representation of floating-point numbers in our numerics). In contrast, above this threshold $\ell_\Delta$ no longer decreases with $L$ (either saturates or gets closer to $0$), and so there is no $\pi$-spectral pairing.

\begin{figure}
    \includegraphics[width=\linewidth]{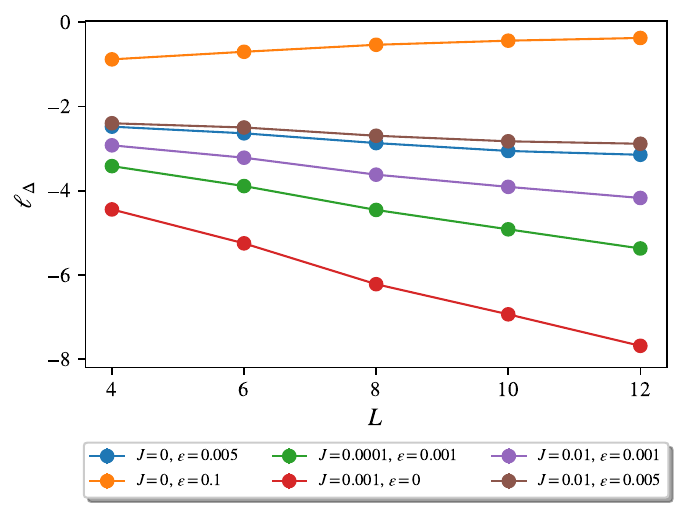}
    \caption{Spectral-paring parameter $\ell_\Delta$ versus $L$ in the spin-1/2 case for different values of $J$, $\varepsilon$ and $\alpha=0.5$. Numerical parameters: $N_d$ varies between $\sim 10^2$ and $\sim 10^4$.}
    \label{fig:Spectral_Pairing_Difference_Log_Delta_non_zero_J_epsilon}
\end{figure}

{In order to further understand the stability of the time crystal phase, we also measure the spectral pairing for different non-zero values of $J$ and $\varepsilon$ in the spin-1/2 case, as seen in Fig.~\ref{fig:Spectral_Pairing_Difference_Log_Delta_non_zero_J_epsilon}. We find again that even with both $J$ and $\varepsilon$ non-zero $\ell_\Delta$ decreases exponentially in system size, albeit this is possible for fairly small values of the perturbations.}

So, we have $\pi$-spectral pairing not only at the solvable point $J=0$, but also in a full range around this value. Therefore, there is a phase where time-translation symmetry breaking appears, as we theoretically expected. Quite nicely, the range of parameters of $J$ -- at $\varepsilon = 0$ -- where $\pi$-spectral pairing appears coincides with the range where the system has a Poisson level statistics (see Fig.~\ref{fig:Gap_Ratio}), and so the dynamics is non-thermalizing, and persistent period-doubling oscillations can appear.

\subsection{Correlations in the Floquet states}\label{amanta:sec}
Let us discuss the structure of the Floquet states. As we have seen above, the majority of these states are the superposition of two different classical spin configurations (cat states) {similar to Eq.~\eqref{cat:eqn}}. The same occurs in the case of the usual spin-flip time crystal~\cite{Potter2017,Nayak2016}. In that case all the Floquet states are cat states, and all them show long-range correlations. Defining $C_\beta(i,j)\equiv\braket{\psi_\beta|\hsigma_i^z\hsigma_j^z|\psi_\beta}-\braket{\psi_\beta|\hsigma_i^z|\psi_\beta}\braket{\psi_\beta|\hsigma_j^z|\psi_\beta}$, one has $C_\beta(1,L)\neq 0$, whatever the state $\ket{\psi_\beta}$ and the length $L$ chosen.

%
In our case the situation is different. Here there can be spin pairs left unchanged by the swap, and the cat-state features might not be fully captured by looking at the correlations between first and last site.
For instance, consider the state which is ferromagnetic in the first half of the chain and antiferromagnetic in the second half, {and construct the Floquet states according to Eq.~\eqref{cat:eqn}} $\ket{\psi_\beta}\propto\ket{++,++,\cdots,+-,+-} + e^{\i\phi}\ket{\text{swap}}$. If we measure correlations only in the first half, or between the first and second half, we would incorrectly classify this as a short-range correlated state, whereas within the second half of the chain there are indeed long-range correlations. Therefore, to properly measure correlations in such state it is necessary to take into account a global {correlation quantifier. We use} the sum of all two-point connected correlation function, divided by the number of spins
\begin{equation}\label{eq:Definition_Sigma_Correlations}
    \Sigma_\beta \equiv \frac{1}{L}\sum_{i<j} \abs{C_\beta(i,j)}\,,
\end{equation}
with the two-site correlation $C_\beta(i,j)$ defined above. Such a quantity measures the amount of correlations per site. Heuristically, we expect two distinct behaviors for short-range and long-range correlated states: for short-range “physical” states, a given site will be correlated only with a small (non-extensive) amount of neighbors, meaning $\Sigma_\beta \sim O(1)$; for long-range “unphysical” states, a given site will be potentially correlated with a finite fraction of the entire chain, meaning that $\Sigma_\beta \sim O(L)$.

Let us first focus on the solvable case. Looking at the definition of the local imbalance Eq.~\eqref{eq:Def_Local_Imbalance} we see that it is kept unchanged under swap, so it has a well-defined value over the Floquet states (that are superposition of two classical configuration related by the swapping operation). With a bit of work (see Appendix~\ref{cocorr:sec}) we can find for $s=1/2$ the simple relation 
 \begin{equation}\label{correli:eqn}
   \Sigma_\beta = {\LI}_{,\beta} ({\LI}_{,\beta}\cdot L-1)/2\,.
 \end{equation}
From it one can see that $\Sigma_\beta$ linearly increases with $L$ if and only if the local imbalance is nonvanishing. This shows why the local imbalance is a natural measure of the time-translation symmetry breaking properties of a Floquet state: It quantifies the amount of long-range correlations related to its cat-state nature. In particular, notice that when $\LI=0$, the eigenstates are trivial spin configurations $\ket{\psi_\beta} = \ket{\{s_k\}}$, where there are no long-range correlations and no spectral pairing. Eq.~\eqref{correli:eqn} is valid only for $s=1/2$. For generic $s$ a similar relation is valid if correlations are evaluated using the mutual information (see Appendix~\ref{mutu:sec}) instead of the connected correlator.

\begin{figure}
    \centering
    \begin{tabular}{c}
         \begin{overpic}[width=80mm]{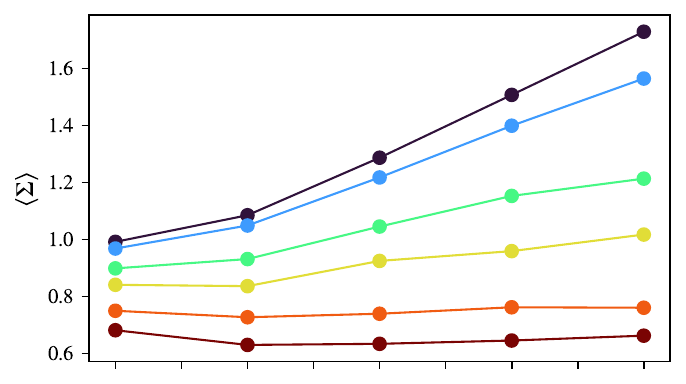}\put(0,48){(a)}\end{overpic}\vspace{-0.2cm}\\
         \begin{overpic}[width=80mm]{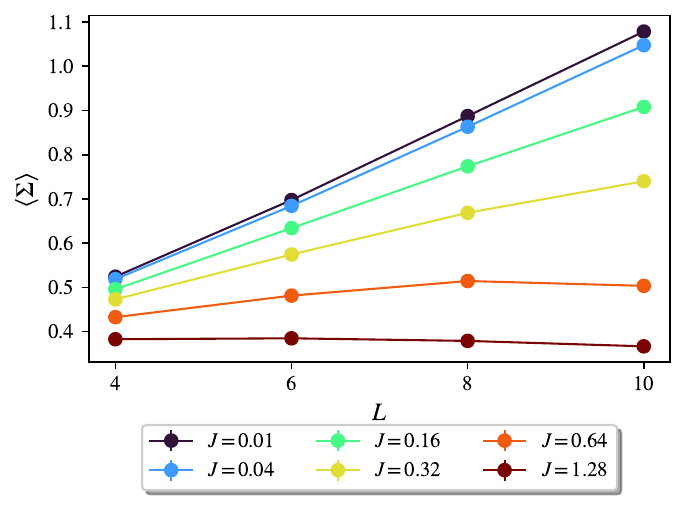}\put(0,65){(b)}\end{overpic}\\
    \end{tabular}
    \caption{$\braket{\Sigma}$ vs $L$ for different values of $J$ in the spin-1/2 (panel a) and spin-1 (panel b) cases. Numerical parameters: $\alpha=0.5$, $N_d = 5120\cdot 2^{1-L/2}$.}
    \label{fig:CORR_avg_wrt_L}
\end{figure}

Let us move to the numerical results in the presence of perturbations. Specifically, we look at the average $\braket{\Sigma}$, defined as
$$
  \braket{\Sigma} = \frac{1}{\calN} \overline{\sum_\beta \Sigma_\beta}\,,
$$
with $\beta$ restricted the $\mathcal{S}_z=0$ subspace. We plot $\braket{\Sigma}$ versus $L$ in Fig.~\ref{fig:CORR_avg_wrt_L}(a). We see two distinct behaviors of $\Sigma_\beta$. At small $J$, close to the solvable point, it increases linearly with $L$, as a signature of long-range correlations, {as we have described above}. {The range of $J$ where this fact occurs is in agreement with the range where spectral pairing occurs [see Fig.~\ref{fig:Spectral_Pairing_Difference_Log_Delta}(a)] and the dynamics is regular [see Fig.~\ref{fig:Gap_Ratio}(a)]. At large $J$ instead $\braket{\Sigma}$ gets smaller and possibly saturates with the system size, and accordingly there is no spectral pairing and no regular dynamics.}

A similar analysis can be performed in the spin-1 case replacing $\hsigma_i^z$ with $\hS_i^z$ in the definition of $C_\beta(i,j)$ and computing the resulting $\Sigma_\beta$ over the Floquet states. At the solvable point no straightforward relation may be found between this quantity and local imbalance. {Nevertheless, when we apply the perturbation $J\neq 0$, we still find that $\braket{\Sigma}$ linearly increases with $L$ in a range of $J$ [see Fig.~\ref{fig:CORR_avg_wrt_L}(b)], marking long-range correlations. In this range there is also spectral pairing [see Fig.~\ref{fig:Spectral_Pairing_Difference_Log_Delta}(b)] and regular dynamics [see Fig.~\ref{fig:Gap_Ratio}(b)]. For larger $J$ there is none of these properties and no time crystal.}~\cite{notagarg} 


%
\section{Conclusions}\label{sec:conclusions}
In conclusion we have described a mechanism to realize time crystals in Floquet systems, based on a spin Hamiltonian with disorder and long-range interactions, to which a time-periodic swapping of nearby sites is applied. In contrast with the existing examples, based on the flipping of the local spins, this protocol is not restricted to spin-1/2 Hamiltonian, but can be applied to models with local spins of any magnitude.

After that, we have introduced the model Hamiltonians, and have analytically discussed the case of the solvable point, where the period-doubling oscillations of the local magnetizations persist forever, also at finite system size. If one takes as initial state a classical spin configuration, one can see period-doubling oscillations for the large majority of these configurations, {with the exception of} a fraction exponentially decaying to zero with system size. In order to characterize which configurations give rise to a period-doubling dynamics, we have introduced a quantity called local imbalance ($\LI$). We have seen that, for increasing local spin magnitude $s$, the spin configurations have in average a larger $\LI$, with a smaller variance. This implies that for increasing system
size a random initial spin configuration shows an increasing tendency to provide a period-doubling behavior

Then we have then moved away from the solvable point, to see if the period-doubling oscillations are robust to such a perturbation. This is an important point to understand if we are dealing with a time crystal, {\em i.\,e.} a robust phase of the matter stabilized by the interactions over a finite parameter range. For that purpose we have numerically studied the model with $s=1/2$ (spin-1/2 case) and the one with $s=1$ (spin-1 case). For perturbations small enough but finite we see that the period-doubling oscillations persist for a time that exponentially increases with the system size: Only in the large-size limit they persist forever. In this way the time-translation symmetry breaking occurs only in the thermodynamic limit, in analogy with the standard symmetry breaking of quantum phase transitions.

{We have then focused on the properties of Floquet quasienergies and states.} We have considered a probe of quantum regularity/ergodicity, the average level spacing ratio of Floquet quasienergies, to understand in which range of parameters the quantum dynamics is regular, thermalization of local observables is absent, and then period-doubling oscillations are possible. We have found a range of parameters where this ergodicity breaking occurs, and the effect becomes more marked when the system size is increased, so that we deal with a regular-dynamics phase. In this range of parameters a nontrivial period-doubling dynamics is possible, and the ergodicity breaking is an effect of disorder that induces many-body localization.

{We have moved to period-doubling related properties of Floquet quasienergies.} We have analytically shown that at the solvable point, up to a fraction exponentially vanishing with the system size, quasienergies show $\pi$-spectral pairing, a property strictly associated to period doubling. Moving away from the solvable point, we have numerically  shown that the $\pi$-spectral pairing is robust in a well-defined range of parameters, that defines the time-crystal phase. This robustness is related to disorder and long-range interactions, that break degeneracies of the quasienergies at the solvable point, making properties of quasienergies and Floquet states robust under local perturbations. The range of parameters where $\pi$-spectral pairing occurs lies inside the range where the dynamics is regular and there is no thermalization.

Looking at the properties of the Floquet states, we have defined a global correlation quantifier. At the solvable point we have found that it is strictly related to the $\LI$ of the Floquet state: a finite $\LI$ corresponds to a Floquet state with the cat-state structure related to $\pi$-spectral pairing. 
Moving away from the solvable point, we have found that there is a finite range of parameters where the correlations increase linearly with system size, which indicates cat-state behaviour and symmetry-breaking. 

{The ranges of parameters where there is regular dynamics, $\pi$-spectral pairing and long-range correlations agree with each other. Together with the results about the dynamics, this suggests the existence of a time-crystal phase.}

{About experimental implementation, at least for the spin-1/2 case, semiconductor spin qubits are a good platform, better than for the spin-flip case. In this context indeed there is time control on the Heisenberg interactions required to implement the swap, while the magnetic field (needed to implement the spin flip) is static~\cite{RevModPhys.95.025003}. The experimental implementation of the partial swap -- that is our swap far from the solvable point -- is discussed in detail in~\cite{nat}.}

Perspectives of future work include the application of this protocol to models with different local Hilbert space (for instance bosons or rotors), and the use of different physical effects to induce ergodicity-breaking (for instance the Stark many-body localization~\cite{PhysRevLett.130.120403,Schulz_2019}).


\subsection*{Acknowledgements}

We thank M.~Fava and R.~Fazio for insightful comments on the manuscript. We acknowledge the Peter-Gr\"unberg Institute, Quantum Control (PGI-8), Forschungszentrum J\"ulich for providing computational resources that facilitated the completion of this work. G.~P. and A.~R. acknowledge financial support from PNRR MUR project PE0000023-NQSTI. 
P.L. acknowledges computational resources from MUR, PON “Ricerca e Innovazione 2014-2020”,  under Grant No. PIR01\_00011 - (I.Bi.S.Co.).
P.L. acknowledges financial support from PNRR
MUR project CN\_00000013-ICSC, as well as from the project QuantERA II Programme STAQS project that has received funding from the European Union’s Horizon 2020 research and innovation programme under Grant Agreement No 101017733.
R.~G. acknowledges funding by the Deutsche Forschungsgemeinschaft (DFG, German Research Foundation) under Germany's Excellence Strategy – Cluster of Excellence Matter and Light for Quantum Computing (ML4Q) EXC 2004/1 – 390534769.

\appendix
\section{Spin swapping} \label{swap:sec}
{
Here we show that the operator in Eq.~\eqref{eqn:swap} realizes swaps product states. It suffices to show this in the $\hsigma_z$ basis $\ket{s_i,s_j}$, and then extends the result linearly to all product states. So we wish to show
\begin{equation}
    \frac{1+\vsigma_i\cdot\vsigma_j}{2}\ket{s_i,s_j} = \ket{s_j,s_i}.
\end{equation}
We use the defining identities of the Pauli operators $\hsigma_z\ket{s} = s\ket{s}$, $\hsigma_x\ket{s} = \ket{-s}$ and $\hsigma_y = \i\hsigma_x\hsigma_z$, hence:
\begin{equation}
    \begin{split}
        \vsigma_i\cdot\vsigma_j &= \hsigma_i^z\hsigma_j^z + \hsigma_i^x\hsigma_j^x + \hsigma_i^y\hsigma_j^y =\\
        &= \hsigma_i^z\hsigma_j^z + \hsigma_i^x\hsigma_j^x(1-\hsigma_i^z\hsigma_j^z),\\
        \frac{1+\vsigma_i\cdot\vsigma_j}{2} &= \frac{1+\hsigma_i^z\hsigma_j^z}{2} + \hsigma_i^x\hsigma_j^x\frac{1-\hsigma_i^z\hsigma_j^z}{2}.
    \end{split}
\end{equation}
Since the $\hsigma_k^z$ operators act diagonally on the spin basis, they serve purely as a filter to either the condition $\delta_{s_i,s_j} = \frac{1+s_is_j}{2}$ (first term, aligned spins) or the condition $\delta_{s_i,-s_j} = 1 - \delta_{s_i,s_j} = \frac{1-s_is_j}{2}$ (second term, unaligned spins):
\begin{equation}
    \frac{1+\vsigma_i\cdot\vsigma_j}{2}\ket{s_i,s_j} = \begin{cases}
        \ket{s_i,s_j} & s_i = s_j,\\
        \ket{-s_i,-s_j} & s_i = -s_j,
    \end{cases} = \ket{s_j,s_i},
\end{equation}
as expected. From this relation one can also see how in the spin-1 case the swap operator has to involve higher power of the spin operators, since already the "filter" itself $\delta_{s_i,s_j}$ (and its complement) involve terms such as $s_i^2+s_j^2$, as seen in Eq.~\eqref{eq:Def_Local_Imbalance_Spin-1}.
}

We now aim to demonstrate Eq.~\eqref{swappero:eqn}. This is equivalent to demonstrate the following equality
\begin{equation}
  e^{-\i(\pi/4)\vsigma_{i}\cdot\vsigma_{j}} = e^{\i\phi} \frac{\Id+\vsigma_{i}\cdot\vsigma_{j}}{2}\,,
\end{equation}
with $i$ and $j$ arbitrary sites and some phase $\phi$. To do this, we use the fact that the swap operator is an involution and thus squares to the identity:
\begin{equation}
    \left(\frac{\Id+\vsigma_{i}\cdot\vsigma_{j}}{2}\right)^2 = \Id.
\end{equation}
This allows us to expand the exponential in a power series, which corresponds to Euler's identity
\begin{equation}
    e^{-\i\frac{\pi}{2}\frac{\Id+\vsigma_{i}\cdot\vsigma_{j}}{2}} = -\i \frac{\Id+\vsigma_{i}\cdot\vsigma_{j}}{2}
\end{equation}
and leads to the result we were looking for
\begin{equation}
    e^{-\i\frac{\pi}{4}\vsigma_i\cdot\vsigma_j} = e^{-\i\pi/4}\frac{\Id+\vsigma_{i}\cdot\vsigma_{j}}{2}.
\end{equation}
A similar derivation can also be done for the general spin-$s$ case, by starting with the swap operator $\hSw$, which satisfies $\hSw^2 = \Id$:
\begin{equation}
    e^{-\i\frac{\pi}{2}(\hSw-1)} = \hSw.
\end{equation}
In the case of spin-1 we arrive at the desired result:
\begin{equation}
    e^{-\i\frac{\pi}{2}( (\vS_i\cdot\vS_j)^2 + \vS_i\cdot\vS_j)} = -( (\vS_i\cdot\vS_j)^2 + \vS_i\cdot\vS_j - 1).
\end{equation}
\section{$\pi$ spectral pairing and period doubling}\label{app:pidoub}
Let us show the relation between $\pi$-spectral pairing and period doubling (see also~\cite{Russomanno_PRL12,Surace_2019}). Let us assume that at least a part of the Floquet spectrum is $\pi$-spectral paired, that is to say, it is organized in pairs $\mu_\beta^-$, $\mu_\beta^+$ such that $\mu_\beta^--\mu_\beta^+=\pi$. We initialize the system in the state $\ket{\psi(0)}$ and study the stroboscopic dynamics of some observable $\mathcal{O}(t) = \braket{\psi(t)|\hat{\mathcal{O}}|\psi(t)}$, with $t$ an integer multiple of the driving period $T=1$. We can expand the stroboscopic-time-evolved state in the Floquet basis as $\ket{\psi(t)}=\sum_\beta R_\beta\nep^{-i\mu_\beta t}\ket{\psi_\beta}$, where $R_\beta \equiv \braket{\psi_\beta|\psi(0)}$. Expanding the expression for the expectation of the observable we get
\begin{align}
  \mathcal{O}(t) &= \underbrace{\sum_\beta |R_\beta|^2 \braket{\psi_\beta|\hat{\mathcal{O}}|\psi_\beta}}_{\text{diagonal}}\nonumber\\
     &+ \underbrace{\sum_\beta (R_\beta^+)^*R_\beta^-\braket{\psi_\beta^-|\hat{\mathcal{O}}|\psi_\beta^+}\nep^{i(\mu_\beta^--\mu_\beta^+)t}}_{\text{period-doubling}}\nonumber\\
     &+\underbrace{\sum_{\beta\neq \gamma,\,\nu\neq \rho}(R_\beta^\nu)^*(R_\gamma^\rho) \braket{\psi_\beta^\nu|\hat{\mathcal{O}}|\psi_\gamma^\rho}\nep^{i(\mu_\beta^\nu-\mu_\gamma^\rho)t}}_{\text{off-diagonal}}\,.
\end{align}
The diagonal term provides a constant contribution, while the off-diagonal one (where $\mu_\beta^\nu$ and $\mu_\gamma^\rho$ are different and unpaired) gives a superposition of incoherent oscillations that, after disorder-average, vanish after a short transient. The interesting term is the one denoted ``period-doubling'', where one can factorize a term $\nep^{i\pi t}$ that provides the period-doubling oscillations, due to the fact that $\mu_\beta^--\mu_\beta^+ = \pi$.

We have seen in Sec.~\ref{numerello:sec} that when we are not at the solvable point but $J$ is small enough, the $\pi$-spectral pairing appears only in the thermodynamic limit. At any finite size one has $\mu_\beta^--\mu_\beta^+ = \pi+\delta_\beta$, where $\delta_\beta$ is a small correction decreasing with the system size. Due to this correction, the period-doubling term averaged over disorder provides period-doubling oscillations that last for a finite time, provided roughly by the inverse of the typical value of $\delta_\beta$ (call it $\delta$). Because $\delta$ decreases with the system size (see Sec.~\ref{numerello:sec}), the time of the period-doubling oscillations increases exponentially with the size, as we have seen in Secs.~\ref{spin12_dyn:sec} and~\ref{spin1dyn:sec}.

We emphasize that the period-doubling term is nonvanishing only for an appropriate choice of operator and initial state. For instance, in Fig.~\ref{fig:Half_Néel_Comparison_sigmaz_dynamics}, the period-doubling term vanishes for the onsite magnetizations at the sites corresponding to the half of the chain where the initial state is uniform.

\section{Short-range results ($\alpha=3$)}\label{app:short-range-figures}

\begin{figure}
    \centering
    \begin{overpic}[width=80mm]{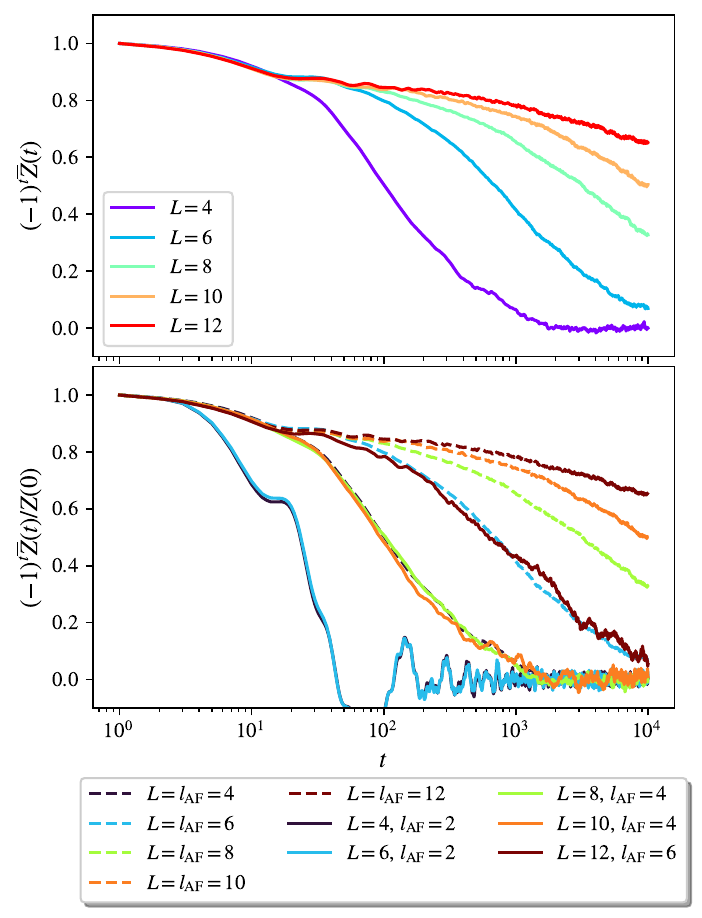}\put(0,54){(b)}\put(0,92){(a)}\end{overpic}
    \caption{Spin-1/2 model. (Panel a) $(-1)^t\overline{Z}(t)$ versus $t$ for initial Néel state and different system sizes. (Panel b) $(-1)^t\overline{Z}(t)$ versus $t$ for  initial Half-Néel state (solid line) and Néel state (dashed line) for different system sizes. Initial values have all been normalized to $+1$ to make the comparison clearer. Parameters: $J=0.1$, $\varepsilon = 0.01$.}
    \label{fig:SpinHalf_Z_dynamics_alpha3}
\end{figure}

\begin{figure}
    \centering
    \begin{overpic}[width=80mm]{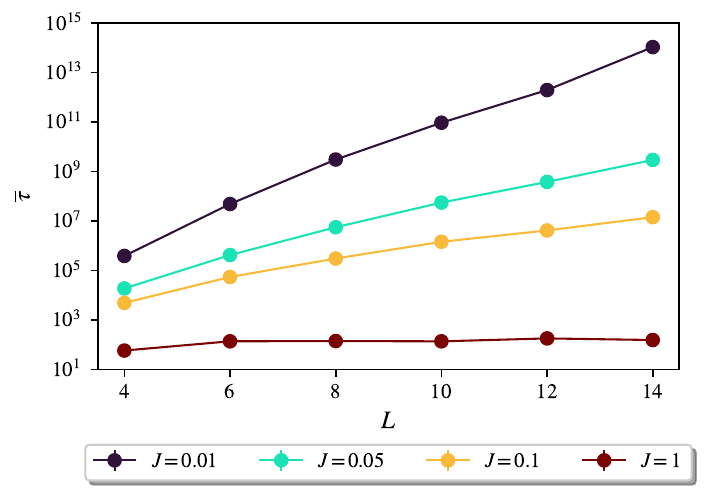}\put(0,52){(a)}\end{overpic}\\
    \begin{overpic}[width=80mm]{Decay_Times_1-2_3}\put(0,52){(b)}\end{overpic}
    \caption{(Panel a) Spin-1/2 model. Decay time $\overline{\tau}$ versus $L$ for initial Néel state and different values of $J$. Here we have considered time evolutions while skipping over $n\geq 1$ periods, such that the time step stays below $\sim 1\%$ of the standard deviation of $\tau$. (Panel b) Spin-1 model. Decay time $\overline{\tau}$ versus $L$ for initial Up-Zero state.}
    \label{fig:Decay_Times_Neel_alpha3}
\end{figure}

\begin{figure}
    \centering
    \begin{tabular}{c}
         \begin{overpic}[width=80mm]{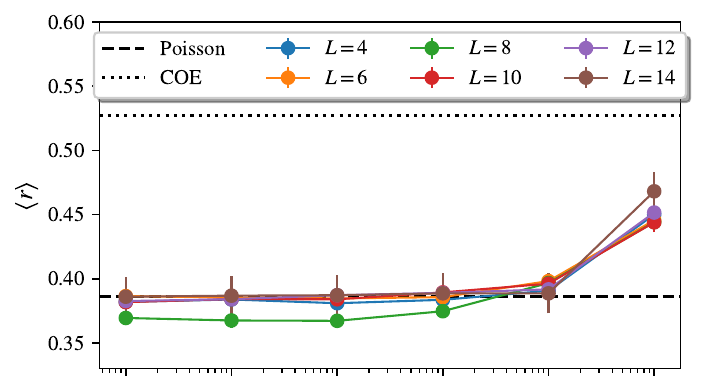}\put(0,52){(a)}\end{overpic}\vspace{-0.3cm}\\
         \begin{overpic}[width=80mm]{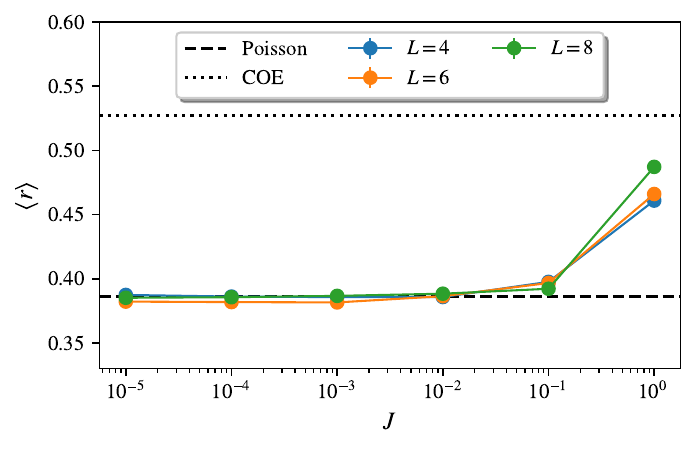}\put(0,52){(b)}\end{overpic} \vspace{-0.5cm}
    \end{tabular}
    \caption{Average level spacing ratio $\braket{r}$ versus $J$ in the spin-1/2 case (panel a) and spin-1 case (panel b) for different $L$ and $\alpha=3$.}
    \label{fig:Gap_Ratio_alpha3}
\end{figure}

\begin{figure}
    \centering
    \begin{tabular}{c}
         \begin{overpic}[width=80mm]{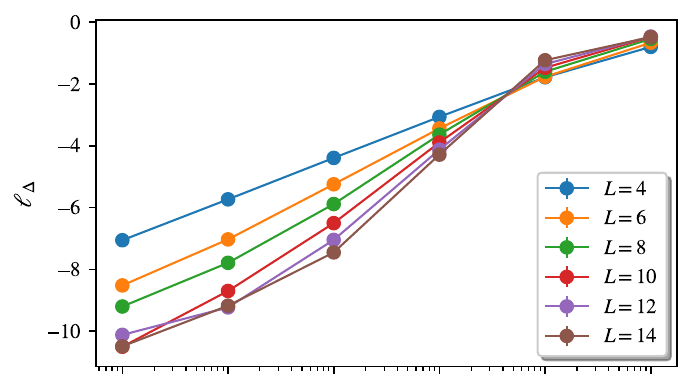}\put(0,52){(a)}\end{overpic}\vspace{-0.3cm}\\
         \begin{overpic}[width=80mm]{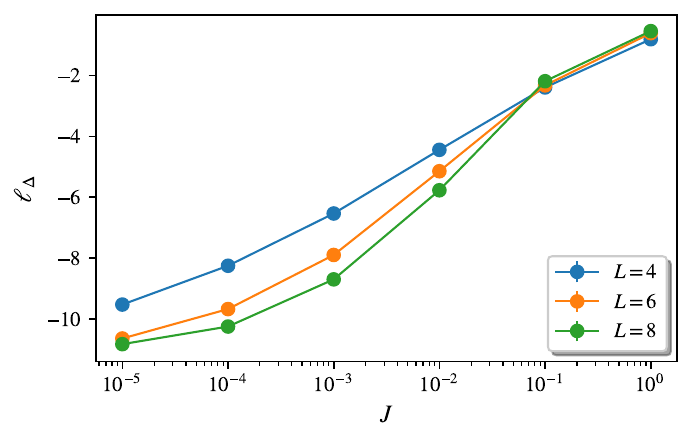}\put(0,52){(b)}\end{overpic} \vspace{-0.5cm}
    \end{tabular}
    \caption{Spectral-pairing parameter $\ell_\Delta$ versus $J$ in the spin-1/2 case (panel a) and spin-1 case (panel b) for different $L$ and $\alpha=3$. }
\label{fig:Spectral_Pairing_Difference_Log_Delta_alpha3}
\end{figure}

\begin{figure}
    \centering
    \begin{tabular}{c}
         \begin{overpic}[width=80mm]{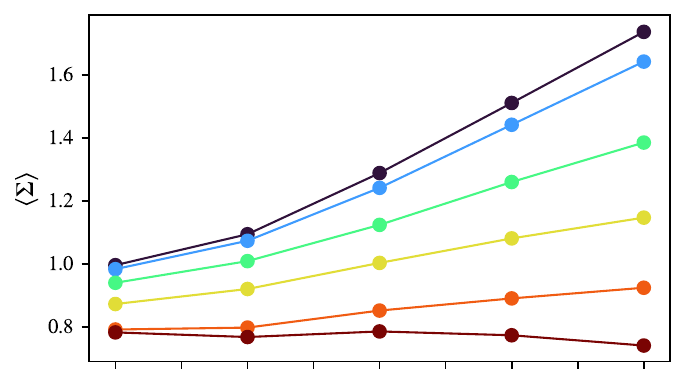}\put(0,52){(a)}\end{overpic}\vspace{-0.2cm}\\
         \begin{overpic}[width=80mm]{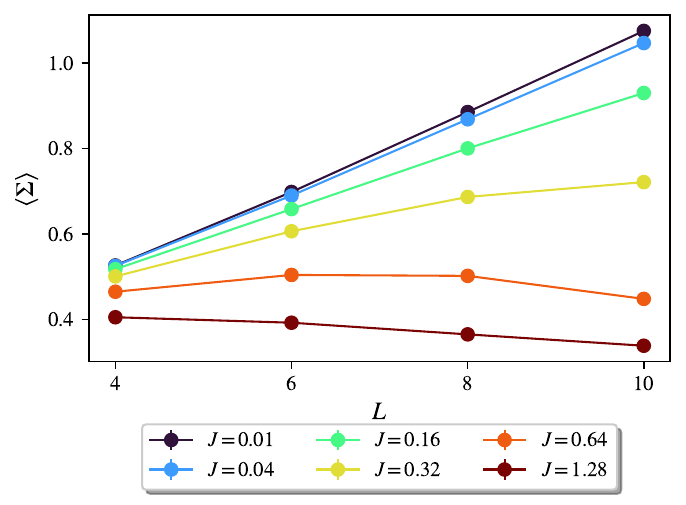}\put(0,52){(b)}\end{overpic}\\
    \end{tabular}
    \caption{$\braket{\Sigma}$ vs $L$ for different values of $J$ in the spin-1/2 (panel a) and spin-1 (panel b) case. Numerical parameters: $\alpha=3$, $N_d = 5120\cdot 2^{1-L/2}$.}
    \label{fig:CORR_avg_wrt_L_alpha3}
\end{figure}

Here we show the results for the short-range case, $\alpha=3$. As mentioned in the main body of the paper, we only found minor quantitative differences between the two. In this section it is taken for granted that $\alpha=3$ in every figure. Let us briefly list the figures:
\begin{itemize}
  \item In Fig.~\ref{fig:SpinHalf_Z_dynamics_alpha3}(a) we show $(-1)^t\overline{Z}(t)$ versus $t$ for different system sizes and initial states in the spin-1/2 model. This is the equivalent of Fig.~\ref{fig:SpinHalf_Z_dynamics}.
  \item In Fig.~\ref{fig:Decay_Times_Neel_alpha3}(a) we show the decay time $\overline{\tau}$ versus $L$ for different parameters in the spin-1/2 model. This is the equivalent of Fig.~\ref{fig:Decay_Times_UpZero_alpha0.5}. In Fig.~\ref{fig:Decay_Times_Neel_alpha3}(b) we show the same the spin-1 model (equivalent of Fig.~\ref{fig:Decay_Times_UpZero_alpha0.5}).
  \item  In Fig.~\ref{fig:Gap_Ratio_alpha3} we show $\braket{r}$ versus $J$ for both models and different system sizes. This is the equivalent of Fig.~\ref{fig:Gap_Ratio}.
  \item In Fig.~\ref{fig:Spectral_Pairing_Difference_Log_Delta_alpha3} we show the spectral-pairing parameter defined in Eq.~\eqref{speppe:eqn} versus $J$ for both models and different $L$. This is the equivalent of Fig.~\ref{fig:Spectral_Pairing_Difference_Log_Delta}.
  \item In Fig.~\ref{fig:CORR_avg_wrt_L_alpha3} we show $\braket{\Sigma}$ vs $L$ for different values of $J$ and both models. This is the equivalent of Fig.~\ref{fig:CORR_avg_wrt_L}.
\end{itemize}
Comparing Figs.~\ref{fig:Gap_Ratio_alpha3},~\ref{fig:Spectral_Pairing_Difference_Log_Delta_alpha3}, and~\ref{fig:CORR_avg_wrt_L_alpha3} we see that also for $\alpha=3$ regular dynamics, spectral pairing and long.range correlations occur in the same range of $J$.

\section{Local imbalance for spin states} \label{lisp:sec}
As discussed in section~\ref{sec:Spin-1_Model}, the spin-1 case displays many differences with the spin-1/2 case. From the point of view of the spectrum, at the solvable point, we can already notice some variations. Apart from the fact that a larger fraction of the spectrum participates to period-doubling at any $L$, a generic spin configuration will also tend to have a larger local imbalance. Quantitatively, this may be seen by looking at the asymptotic behavior $L\to\infty$ of the probability density function $p(\LI)$ of local imbalance over spin configurations. In order to get this density function let us consider 
%
%
the case of a chain of length $L$ where at each site there corresponds a local Hilbert space of dimension $d=2s+1$, where $s$ is the value of the local spin. {It is easy to see that the $\LI$ can only take values
$$
  \LI=\frac{N}{L/2}\,,
$$
where $N\leq L/2$ is the number of pairs whose spins differ from each other. There are many spin configurations associated to a single value of $\LI$. This degeneracy, {\em i.\,e.} the dimension $D_L(\LI)$ of an eigenspace corresponding to a given value of local imbalance $\LI$ is
\begin{equation}
  D_L\left(\LI=\frac{N}{L/2}\right) = (d(d-1))^{N}d^{L/2-N}\binom{L/2}{N}\,.
\end{equation}
One can find this result considering that each pair where spins differ can appear in $d(d-1)$ different ways, while each pair where spins are equal can appear in $d$ different ways. So the probability distribution of the imbalances
$$
  p(\LI) = \frac{1}{d^L}D_L\left(\LI=\frac{N}{L/2}\right)
$$
is a binomial.} 
Therefore, the asymptotic distribution $p(\LI)$ is a normal distribution with mean 
\begin{equation}\label{media:eqn}
  \braket{\LI}=\frac{(d-1)}d
\end{equation}
and variance 
\begin{equation}\label{sigma:eqn}
  \sigma^2(\LI)=p(1-p)/(L/2) = \frac{2(d-1)}{Ld^2}
\end{equation}
vanishing for $L\to\infty$ and $s\to\infty$. This proves more concretely that increasing $s$ improves the behavior of the period-doubling caused by the swap, because the typical value of the local imbalance increases and the fluctuations around it decrease. Although no such direct analytical result can be found restricting to the subspace $\mathcal{S}_z=0$, the same qualitative behavior is observed numerically at large $L$ (not shown). This suggests once again that the subspace $\mathcal{S}_z=0$ we numerically consider here faithfully represents the entire Hilbert space.

\begin{figure}
    \centering
    \includegraphics[width=\linewidth]{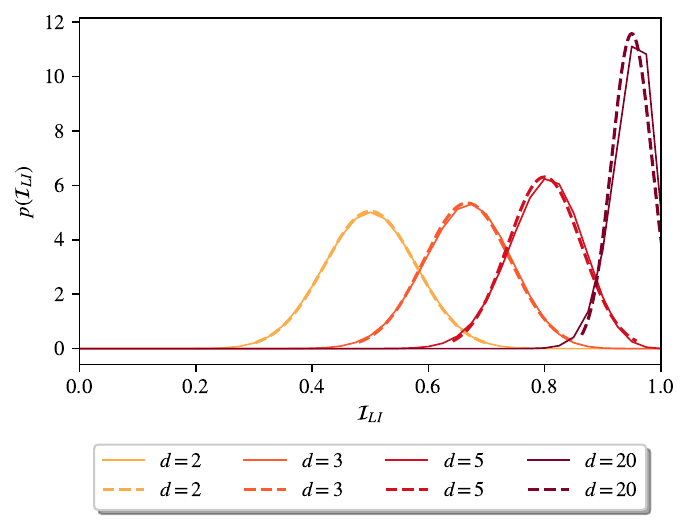}
    \caption{Probability Density Function of the Local Imbalance at different local Hilbert space dimensions $d$ and $L=120$ (solid line) and its approximation as a normal distribution (dashed line). Spin-1/2 and spin-1 correspond to $d=2$ and $d=3$, respectively. }
    \label{fig:Distribution_LI}
\end{figure}


\section{Correlations in the exact Floquet states} \label{corr_exact:app}
In this Appendix we find the relation between the local imbalance and the global correlation quantifier Eq.~\eqref{eq:Definition_Sigma_Correlations}. In Sec.~\ref{cocorr:sec} we consider the case where the global quantifier is constructed using the connected correlations and the relation is valid only for local spin magnitude $s=1/2$. In Sec.~\ref{mutu:sec} we evaluate correlations using the mutual information and get a relation valid for generic $s$.
\subsection{Connected correlations}\label{cocorr:sec}

Here we derive Eq.~\eqref{correli:eqn} relating the local imbalance and the global correlation quantifier Eq.~\eqref{eq:Definition_Sigma_Correlations}. Eq.~\eqref{correli:eqn} is valid in the solvable case ($\varepsilon=0=J$), for the spin-1/2 chain. In this case the Floquet states are given by superpositions of a classical spin configuration and its swapped partner:
\begin{equation}
    \ket{\psi_\beta} = c_+\ket{\{s_k\}} + c_- \ket{\{s_{\overline{k}}\}},
\end{equation}
where $\abs{c_+} = \abs{c_-} = 1/\sqrt{2}$ if $\LI>0$ while $c_+=1$, $c_-=0$ (or equivalently $c_-=1$, $c_+=0$) if $\LI=0$. Since we are dealing with expectation values of $\hsigma^z$ operators and their products, the phases of $c_+$ and $c_-$ are irrelevant and we ignore them for convenience of notation. 

We first notice that the spins at sites $k$ where $s_k = s_{\overline{k}}$ are independent from every other spin, since their value remains the same in the superposition. This means that the state factorizes as a product state between site $k$ and every other site:
\begin{equation}
    \begin{split}
        \ket{\psi_\beta} = c_+&\ket{s_1\cdots s_k \cdots} +
        c_- \ket{s_{\overline{1}}\cdots s_k \cdots} = \ket{\psi_{\beta}}_{\not{k}} \otimes \ket{s_k}\,.
    \end{split}
\end{equation}
Therefore, any connected correlation function between such sites $k$ and any other site $q$ will be zero, $C_\beta(k,q) = 0,\;\forall q\neq k$.

There only remains the case where both $k$ and $q$ satisfy $s_k = -s_{\overline{k}}$, $s_q = -s_{\overline{q}}$. The evaluation of $C_\beta(k,q)$ is now trivial if one notices that $s_ks_q = s_{\overline{k}}s_{\overline{q}}$:
\begin{equation}
    \begin{split}
        \braket{\hsigma_k^z\hsigma_q^z}_\beta &= \abs{c_+}^2\bra{\{s_i\}}\hsigma_k^z\hsigma_q^z\ket{\{s_i\}}
        + \abs{c_-}^2\bra{\{s_{\overline{i}}\}}\hsigma_k^z\hsigma_q^z\ket{\{s_{\overline{i}}\}}\\& = s_ks_q,\\\\
        \braket{\hsigma_k^z}_\beta &= \abs{c_+}^2\bra{\{s_i\}}\hsigma_k^z\ket{\{s_i\}} + \abs{c_-}^2\bra{\{s_{\overline{i}}\}}\hsigma_k^z\ket{\{s_{\overline{i}}\}} \\&= \abs{c_+}^2(s_k - s_{\overline{k}}) = 0,\\\\
        C_\beta(k,q) &= \braket{\hsigma_k^z\hsigma_q^z}_\beta - \braket{\hsigma_k^z}_\beta\braket{\hsigma_q^z}_\beta = s_ks_q.
    \end{split}
\end{equation}
In the spin-1/2 case this can only take values $\pm 1$, meaning that its absolute value is always $+1$. Finally, we have that $\abs{C_\beta(k,q)}$ vanishes for every site $k$ aligned with its partner $\overline{k}$, while it equals one for every choice of sites $k$ and $q$ that are both unaligned with respect to their partners $\overline{k}$ and $\overline{q}$. By definition of local imbalance, the number of such sites is equal to ${\LI}_{,\beta} \cdot L$ (double the number of unaligned pairs). Finally, the computation of $\Sigma_\beta$ reduces to counting the number of distinct pairs of sites within this region:
\begin{equation}
    \begin{split}
        \Sigma_\beta = \frac{1}{L}\sum_{i<j} \abs{C_\beta(i,j)} = \frac{1}{L}\sum_{i|s_i \neq s_{\overline{i}}} &\,\sum_{j>i | s_j \neq s_{\overline{j}}} +1 =\\ 
        = \frac{{\LI}_{,\beta} ({\LI}_{,\beta} \cdot L - 1)}{2},
    \end{split}
\end{equation}
which is the desired result.


\subsection{Mutual information}\label{mutu:sec}
A general, model-independent, alternative to connected correlation functions which can measure correlations between two sites is given by the mutual information $\calI$:
\begin{equation}
    \calI_\beta(i,j) = S_i^{(\beta)} + S_j^{(\beta)} - S_{i,j}^{(\beta)},
\end{equation}
where $S_A^{(\beta)} = -\mathrm{Tr}(\rho_A^{(\beta)}\ln\rho_A^{(\beta)})$ is the entanglement entropy of the Floquet state $\ket{\psi_\beta}$ over the region $A$ (i.e. a set of sites in the chain). Calling $B$ the complement of $A$, the reduced density matrix is defined as $\rho_A^{(\beta)}= \mathrm{Tr}_B[\ket{\psi_\beta}\bra{\psi_\beta}]$. The global correlation quantifier for the mutual information is defined as
\begin{equation}
  \Sigma_\beta(\calI) = \frac{1}{L}\sum_{i<j} \calI_\beta(i,j)\,.
\end{equation}

We now show by using the mutual information instead of a connected correlation function that we get the same scaling law in the general spin-$s$ case.
Let us first explicitly write the reduced density matrix of sites $i$ and $j$ for the Floquet state $\ket{\psi_\beta} = c_+\ket{\{s_k\}} + c_-\ket{\{s_{\overline{k}}\}}$:
\begin{equation}
    \begin{split}
        \rho_{i,j}^{(\beta)} &= \mathrm{Tr}_{\not{i}\not{j}} \ket{\psi_\beta}\bra{\psi_\beta}\\
        &= \abs{c_+}^2\ket{s_i,s_j}\bra{s_i,s_j} + \abs{c_-}^2\ket{s_{\overline{i}}, s_{\overline{j}}}\bra{s_{\overline{i}}, s_{\overline{j}}} +\\
        &+ c_+c_-^* \delta_{\{s_k\}_{\not{i}\not{j}}, \{s_{\overline{k}}\}_{\not{i}\not{j}}} \ket{s_i,s_j}\bra{s_{\overline{i}}, s_{\overline{j}}} + \mathrm{h.c.},
    \end{split}
\end{equation}
where each $s_k$ now takes $d=2s+1$ different possible values in $\{-s, -s+1, \cdots, s-1, s\}$.
If $\{s_k\}_{\not{i}\not{j}} \neq \{s_{\overline{k}}\}_{\not{i}\not{j}}$ - which means that there is at least one site $q$ where $s_q \neq s_{\overline{q}}$ - then the reduced density matrix takes the simple form:
\begin{equation}
    \rho_{i,j}^{(\beta)} = \abs{c_+}^2\ket{s_i,s_j}\bra{s_i,s_j} + \abs{c_-}^2\ket{s_{\overline{i}}, s_{\overline{j}}}\bra{s_{\overline{i}}, s_{\overline{j}}}
\end{equation}
and similarly for the reduced density matrix of site $i$ and site $j$:
\begin{equation}
    \begin{split}
        \rho_i^{(\beta)} &= \abs{c_+}^2\ket{s_i}\bra{s_i} + \abs{c_-}^2\ket{s_{\overline{i}}}\bra{s_{\overline{i}}},\\
        \rho_j^{(\beta)} &= \abs{c_+}^2\ket{s_j}\bra{s_j} + \abs{c_-}^2\ket{s_{\overline{j}}}\bra{s_{\overline{j}}},
    \end{split}
\end{equation}
It is straightforward to see that when $\abs{c_+}^2 = \abs{c_-}^2 = 1/2$ and $s_i\neq s_{\overline{i}},\, s_j\neq s_{\overline{j}}$ the entropy of these states is $\ln 2$ and consequently the mutual information is also $\ln 2$:
\begin{equation}
    \calI_\beta(i,j) = S_i + S_j - S_{ij} = \ln 2.
\end{equation}
Outside of an irrelevant special case (at $\LI = \frac{1}{L/2}$), the scaling of the global correlation quantifier is still respected, as only the pairs with unaligned spins will contribute
\begin{equation}\label{scali:eqn}
    \Sigma_\beta(\calI) = \frac{1}{L}\sum_{i<j} \calI_\beta(i,j) = \frac{{\LI}_{,\beta} ({\LI}_{,\beta} \cdot L - 1)}{2}\ln 2. 
\end{equation}
We notice that for $L\gg 1$ a typical Floquet state $\ket{\psi_\beta}$ has $(d-1)/d\cdot L/2$ pairs of unaligned neighbouring spins, corresponding to $\LI = (d-1)/d$. This typical value coincides with the average Eq.~\eqref{media:eqn}, because the fluctuations Eq.~\eqref{sigma:eqn} vanish for $L\to\infty$. Substituting this typical value in Eq.~\eqref{scali:eqn} we find that the typical amount of global correlations increases with local spin magnitude $s$
\begin{equation}
    \overline{\Sigma(\calI)} \sim  \frac{L}{2} \frac{(d-1)^2}{d^2}\ln 2\,,
\end{equation}
and the fluctuations around this value become smaller, as we can see in Eq.~\eqref{sigma:eqn}.
This finding confirms that at higher values of spin the correlations are stronger and thus the time crystalline behaviour becomes more robust with increasing $s$.

%

\begin{thebibliography}{77}
\bibitem [{not({\natexlab{a}})}]{noter}%
  \BibitemOpen
  \href@noop {} {} \bibinfo {note} {{O}ne can see
  time-translation symmetry breaking in the ground state of quite unphysical
  Hamiltonians, made as sums of Pauli strings that have extensive length and
  encompass halves of a spin
  {system~\cite{PhysRevLett.123.210602}}}\BibitemShut {NoStop}%
\bibitem [{not({\natexlab{a}})}]{note:log}%
  \BibitemOpen
  \href@noop {} {} \bibinfo {note} {{T}he use of
  log-averages is here used to better represent the typical value of the
  spectral pairing, rather than directly looking at its average value, since,
  due to disorder, various orders of magnitude will compete with each other
  \cite{PhysRevB.101.024204,RevModPhys.58.801}.}\BibitemShut {Stop}%
\bibitem [{not({\natexlab{b}})}]{notagarg}%
  \BibitemOpen
  \href@noop {} {} \bibinfo {note} {{W}e notice that the
  total amount of correlations in the spin-1 case, as measured by
  $\Sigma_\beta$, is smaller than the spin-1/2 case for the corresponding
  parameters. This may be explained by noticing that in this case there are
  other correlations which we have not taken into account, namely, the ones
  between $(\hS_k^z)^2$ at different sites and the ones between $\hS_k^z$ and
  $(\hS_q^z)^2$. One could use a model-independent quantity to avoid this issue
  such as the two-site mutual information $I(i,j) = S_i + S_j - S_{ij}$, where
  $S_A$ is the entanglement entropy of region $A$. Such a quantity upper bounds
  all normalized connected correlation functions \cite{PhysRevLett.100.070502}, thus
  including all possible such contributions. Furthermore, one can show (see
  Appendix~\ref{mutu:sec}) that, even in the spin-$s$ case, at the
  solvable point, the average mutual information per site has a similar
  behavior as $\Sigma_\beta$ for the solvable point in the spin-1/2 case. Since
  on average the local imbalance increases with spin-$s$, this confirms that
  increasing $s$ results in more correlated Floquet states and a stronger time
  crystal phase.}\BibitemShut {Stop}%
\end{thebibliography}
%
%
\end{document}